# Anti-CD44-conjugated olive oil liquid nanocapsules for targeting pancreatic cancer stem cells


Saúl A. Navarro-Marchal,[1,2,3,4] Carmen Griñán-Lisón,[1,2,3] José M. Entrena,[5,6] Gloria Ruiz-Alcalá,[1,2,3] María Tristán-Manzano,[7] Francisco Martin,[7] Ignacio Pérez-Victoria,[8] José Manuel Peula-García,[9,10,*] and Juan Antonio Marchal,[1,2,3,11,*]

[1]Biopathologyand Regenerative Medicine Institute (IBIMER), Centre for Biomedical Research (CIBM), University of Granada, Granada, 18100, Spain;

[2]Instituto de Investigación Biosanitaria ibs.GRANADA, Universityof Granada, Granada, 18071, Spain.

[3]Excellence Research Unit "Modeling Nature" (MNat), University of Granada, Granada18071, Spain.

[4]Department of Applied Physics, Faculty of Sciences, University of Granada, Granada, 18071, Spain

[5]InstituteofNeuroscience, BiomedicalResearch Center, Universityof Granada, Parque Tecnológico de Ciencias de la Salud, 18100 Armilla, Granada, Spain;

[6]AnimalBehaviorResearchUnit, ScientificInstrumentation Center, Universityof Granada, Parque Tecnológico de Ciencias de la Salud, 18100 Armilla, Granada, Spain.

[7]Genomic Medicine Department, GENYO, Centre forGenomics and OncologicalResearch, Pfizer-Universityof Granada-Andalusian Regional Government, Parque Tecnológico Ciencias de la Salud, Granada, Spain.

[8]Fundación MEDINA, Centro de Excelencia en Investigación de Medicamentos Innovadores en Andalucía, Parque Tecnológico de Ciencias de la Salud, Avda. del Conocimiento 34, 18016 Armilla, Granada, Spain.

[9]Biocolloids and Fluids Physics Group, Faculty of Sciences, University of Granada, 18014 Granada (Spain).

[10]Department of Applied Physics II, University of Málaga, 29071 Málaga, Spain

[11]Department of Human Anatomy and Embryology, Faculty of Medicine, University of Granada, 18016 Granada, Spain.

*Corresponding authors.Prof. Juan Antonio Marchal, MD, PhD; e-mail: jmarchal@ugr.es and Dr. José Manuel Peula-García, PhD; email: jmpeula@uma.es












## Abstract


The latest trends in cancer research and nanomedicine focus on using nanocarriers to target cancer stem cells (CSCs). Specifically, lipid liquid nanocapsules (LLNC) are usually developed as nanocarriers for lipophilic drug delivery. Here, we developed olive oil liquid nanocapsules (O$^2$LNC) functionalized by covalent coupling of an anti-CD44-FITC antibody (αCD44). Firstly, O$^2$LNCs are formed by a core of olive oil were surrounded by a shell containing phospholipids, a non-ionic surfactant and deoxycholic acid molecules. Then, O$^2$LNCs were coated with an αCD44 antibody (αCD44-O$^2$LNC). The optimization of an αCD44 coating procedure, a complete physico-chemical characterization, as well as clear evidence of their efficacy *in vitro* and *in vivo,* were demonstrated. Our results indicate the high targeted uptake of these αCD44-O$^2$LNCs and the increased antitumor efficacy (up to four times) of paclitaxel-loaded-αCD44-O$^2$LNC compared to free paclitaxel in PCSCs. Also, αCD44-O$^2$LNCs were able to selectively target PCSCs in an orthotopic xenotransplant *in vivo* model.


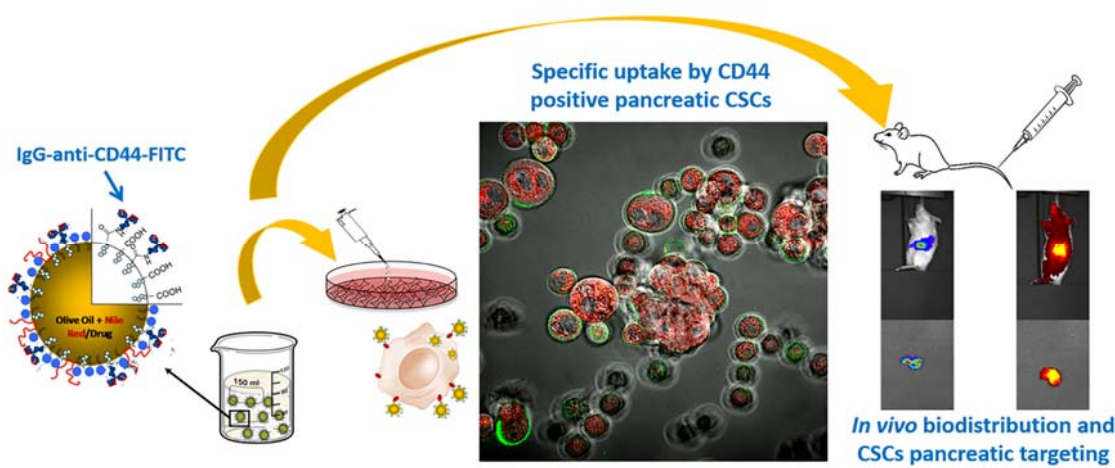

## Graphical abstract





## Introduction

The development of targeted therapies, especially for cancer, is one of the main goals of nanomedicine today[1−3]. Conventional chemotherapy usually prompts a modest tumor response and provokes undesirable side effects because of the nonspecific action of drugs on proliferating tissues[4]. To avoid these and other disadvantages, drug nanocarriers should be formulated to deliver the antitumor drug directly to the cancerous cells. This is a complex interdisciplinary task with too many variables to be properly controlled. These variables include the use of biocompatible materials, with simple but robust processes for biomaterial assembly, usually requiring different conjugation chemistries followed by some purification processes.5 Therefore, current formulations based on complex nanostructures such as polymer conjugates, polymeric micelles, liposomes, carbon nanotubes, or nanoparticles[6−9] must be superficially modified to provide carriers with vectorization properties. In recent years, lipid liquid NCs (LLNCs) have been developed as potential nanocarriers. The inner hydrophobic domain, encased in a hydrophilic outer shell, has been used to encapsulate hydrophobic drugs that are protected during their transport to the target cells[10]. Additionally, the external shell can be physicochemically modified by ligands such as antibodies to deliver the given drug specifically inside the cells that overexpress the corresponding antigen receptor. These tailor-made vehicles could improve the effectiveness of the drug, while minimizing undesirable side effects in healthy cells[11,12]. With regard to the nature of the ligands, monoclonal antibodies have been extensively employed to develop targeted nanocarriers, referred to as immuno-nanocarriers[13−18]. Some advantages of immuno-nanocarriers include the improvement of drug selectivity, high drug-loading efficiency, minor drug leakage, and protection of drugs from enzymatic





degradation[19] .The design of immuno-NCs requires not only the creation of well-defined structures but also the optimization of appropriate antibody-conjugation chemistry.

Cancer-associated aggressiveness and mortality are primarily caused by cancer recurrence and metastasis. Recent advances in cancer research indicate that this is mainly because of a small subpopulation of cells known as cancer stem cells (CSCs or tumor initiating cells), which possess stem-like functional properties such as a self-renewal ability and multipotency and have the ability to differentiate into bulk tumor cells. Although current treatments can reduce the size or eliminate the tumor, however, in several cases, these effects are transient and generally do not improve patient survival outcomes. In fact, during chemotherapy, most of the cells of the primary tumor can be destroyed, but if CSCs are not totally eradicated, they can lead to the reappearance of the tumor in the same organ or in metastatic sites[20]. Therapies that specifically target CSCs hold great promise for improving survival rates and quality of life for cancer patients.

On the other hand, CD44 has attracted considerable attention because of its important functions in mediating cell−cell and cell−matrix interactions and association with malignant processes, particularly in cancer dissemination[21,22]. A recent work has revealed that CD44 is the most common CSC surface marker and plays a pivotal role in CSCs in communicating with the microenvironment and regulating CSC stemness properties[23]. Increasing evidence suggests that CD44 is a promising prognostic biomarker and therapeutic target for many cancers. In fact, it has been demonstrated that populations enriched in CSCs of prostate cancer, uterine adenocarcinoma, and pancreatic cancer (PC) among others have a high expression of CD44[24]. In 2018, there were 460,000 new cases of PC diagnosed around the world[25]. Although there are different types of treatment available for patients with PC, most patients have no





recognizable symptoms, making early diagnosis difficult and chemotherapies or radiation therapies often show poor results[26]. The biochemical and physiological characteristics of PC appear to limit the effectiveness of these standard forms of therapy, mainly due to the highly enriched CSC subpopulations present in this tumor that explains the resistance to treatments[27,28].

In the present study, we have formulated and optimized a new immuno-nanosystem by covalent linking of a specific immuno-γ-globulin (IgG) to the surface of olive oil lipid NC ($O^2$LNC) via carboxylic acid functional groups. Thus, our group has developed a novel system in which the NC's shell was enriched with deoxycholic acid (DCA)[26,27]. Chemically conjugated with oligosaccharides has shown great potential to form both drug and gene carriers[30], and it increases the efficacy of oral bioavailability of different nanoparticles that encapsulate drugs[31]. In this way, using DCA-NC ($O^2$LNC), we achieved the optimal experimental conditions to covalently link the anti-CD44-fluorescein isothiocyanate (FITC) onto their carboxylic acid-rich surfaces, by means of a reproducible and simple method based on a carbodiimide catalyst[32,33]. The antibody is a recombinant humanized IgG1 that specifically recognizes the CD44 antigen overexpressed on the surface of pancreatic CSCs (PCSCs). Subsequently, the obtained olive oil immuno-NCs (αCD44-$O^2$LNC) were physicochemically characterized.

Moreover, both bare $O^2$LNC and αCD44-$O^2$LNC labeled with a fluorescent molecule (nile red or IR-780 iodide) were evaluated *in vitro*, in differentiated PC cells (PCCs) growing in monolayer (low expression of CD44) and PCSCs (high expression of CD44); and *in vivo* analyzing the biodistribution and targeting properties of αCD44-$O^2$LNC in a PCSC orthotopic xenotransplant *in vivo* model.





## Experimental Section

**Reagents.** Poloxamer 188 (Pluronic F-68), deoxycholic acid (DC), Nile Red (RN), IR-780 iodide (i780), N-(3-Dimethylaminopropyl)-N′-ethylcarbodiimide hydrochloride (ECDI), β-Mercapto-ethanol, Brilliant Blue R250, Paclitaxel (PTX) with a purity ≥97.0% and olive oil were purchased from Sigma-Aldrich (Spain). All of them, except the olive oil, were used as received. Olive oil was previously purified in our laboratories with activated magnesium silicate (Florisil, Fluka) to eliminate free fatty acids. Epikuron® 145 V, which is a phosphatidylcholine-enriched fraction of soybean lecithin, was supplied by Cargill (Barcelona, Spain). Human anti-CD44-FITC (αCD44) (clone REA690), anti-CD44-APC (clone REA690), anti-CXCR4-APC andanti-CXCR4-PE (clone REA649) monoclonal antibodies were obtained from Miltenyi Biotech (Spain). Their isoelectric point (IEP) was 6.8 ± 0.5. 3-(4,5-Dimethylthiazol-2-yl)-2,5-diphenol tetrazolium bromide (MTT) cell-proliferation kit was obtained from Promega (U.S.A.), Cell Counting Kit-8 (CCK8) cell-proliferation assay was purchased from Dojindo (Europe). Water was purified in a Milli-Q Academic Millipore system. NMR solvents ($D_2O$, $CDCl_3$ and $CD_3OD$) were purchased from Eurisotop (France). Other solvents and chemicals used were of the highest grade commercially available.

**Cell Lines and Culture Conditions**. 293T human kidney cancer cell line (CRL11268) andBxPC-3human pancreatic cancer cell line was obtained from American Type Culture Collection (ATCC), and cultured following ATCC recommendations. 293T cell line was cultured in DMEM (Biowest) and BxPC-3 pancreatic cancer (PC) cell line was cultured in RPMI-1640 Medium (RPMI) (Gibco, Grand Island, NY, USA) supplemented with 10% (v/v) heat-inactivated fetal bovine serum (FBS) (Gibco) 1% L-glutamine, 2.7% sodium bicarbonate, 1% Hepes buffer, and 1% penicillin/streptomycin





solution (GPS, Sigma). Both cell lines were grown at 37 °C in an atmosphere containing 5% $CO_2$.

Cancer stem-like cells from BxPC3 PC (adenocarcinoma) cell line were isolated using the methodology previously described by us[34]. Briefly, the BxPC-3 cell line was cultured into Ultra-Low Attachment 6-well plates (Corning) with DMEM-F12 (Sigma-Aldrich), 1% streptomycin-penicillin (Sigma-Aldrich), 1 mg/mL hydrocortisone (Sigma-Aldrich), 4 ng/mL heparin (Sigma-Aldrich), 1X ITS (Gibco); 1X B27 (Gibco), 10 ng/mL EGF (Sigma-Aldrich), 10 ng/mL FGF (Sigma-Aldrich), 10 ng/mL of IL-6 (Sigma-Aldrich) and 10 ng/mL of HGF (Sigma-Aldrich). To obtain the second generation of PCSCs used in the experiments, spheres were collected by centrifugation at 1500 rpm for 5 min, incubated 5 min at 37 °C with trypsin-EDTA (Sigma-Aldrich), disaggregated with a 30G needle and washed with 1X PBS. The cells were seeded again in the same type plates and with the same medium mentioned above. After three days, the spheres were collected and used in the different trials[34]. All cell cultures tested were negative for mycoplasma infection.

**Preparation of Olive Oil Lipid Nanocapsules ($O^2$LNC).** $O^2$LNC were prepared by using a modified solvent-displacement technique of Calvo et al.[35] Briefly, an organic phase composed of 125 µL of olive oil, 10 mg of DCA, 40 mg of Epikuron 145 V dissolved in 1 mL of ethanol, and 9 mL of acetone was strongly added to 20 mL of aqueous phase containing 50 mg of Pluronic F68 under magnetic stirring. The mixture turned milky immediately because of the formation of a nanoemulsion. Organic solvents (acetone and ethanol) plus a portion of the volume of water were evaporated in a rotary evaporator at 35 °C, giving a final volume of 16 mL. In some cases, we dissolved PTX or/and NR or i780 iodide dye in the olive oil phase in order to synthesize





$O^2LNC$ loaded with these compounds inside[33]. These different $O^2LNCs$ were used depending on the experiment carried out.

**NMR chemical and structural characterization of the Nanocapsules.** NMR spectra were recorded on a Bruker Avance III spectrometer (500 MHz for $^1H$) equipped with a 1.7mm MicroCryoprobe, using external acetone referencing for the analysis of intact nanocapsules in $H_2O$-$D_2O$ or the signal of the residual solvent as internal reference ($\delta_H$ 7.26 for $CDCl_3$, $\delta_H$ 3.31 for $CD_3OD$) for the analysis of disintegrated nanocapsules reconstituted in these deuterated organic solvents. For nanocapsule disintegration, a clean aliquot (80 µL) of sample was dissolved in 2-propanol (2 mL), mixed, sonicated for 20 minutes and split into two further aliquots which were concentrated to dryness before reconstitution in organic deuterated solvent (ca. 50 µL of $CDCl_3$ or $CD_3OD$) for NMR analyses. The diffusion NMR experiments were undertaken after the size of the nanocapsules had been already determined by dynamic light scattering [photon correlation spectroscopy (PCS)].

**Preparation of Antibody Coated Nanocapsules.** Antibodies were immobilized on the $O^2LNC$ surface by an optimized carbodiimide method previously described[29], which permits an efficient covalent binding of protein molecules to the carboxylic groups supplied by deoxycholic molecules. The antibody-coupling protocol was conducted at pH value, depending on the IEP of the IgG, in such a way that the $O^2LNCs$ were first centrifuged in a Vivaspin 20 centrifugal concentrator MWCO 1000 kDa (Sartorius) at 2500xG for 15 min, this nanofiltration step removed any 'smaller molecular component' from the sample and eliminate the non-reactive products coming from the synthesis protocol. We replaced the medium with a phosphate buffer (PB pH 7.4) for αCD44-FITC. Once the centrifugation was finished, 1 mg of a solution in phosphate buffer (pH 7.4) of ECDI at 15 mg/ml was added to the $O^2LNC$ solution having a total





particle surface equal to 0.29 m$^2$ for the covalent binding of αCD44-FITC (αCD44). Subsequently, the antibody coverage was performed by adding αCD44 concentration of 0.18 mg/m, and then the solution was incubated at room temperature for 2 h. Finally, the αCD44-O$^2$LNCs were centrifuged in the Vivaspin tubes with its corresponding buffer to remove any IgG molecules that were not coupled to the O$^2$LNC and collecting the first elution volume for protein quantification by ultraviolet spectrophotometry and bicinchoninic acid assay methods. The final αCD44-O2LNC stock was stored at 4 °C for further use.

**Protein Separation by SDS-PAGE.** In order to verify that the covalence of the antibody was effective, a gel chromatography test was performed. The immune-nanocapsule complexes were separated and denatured by boiling 20 µl of each sample for 5 min in the following buffer: 62.5 mM Tris-HCl (pH 6.8 at 25 °C), 2% (w/v) sodium dodecyl sulfate (SDS), 10% glycerol, 0.01% (w/v) bromophenol blue, and 40 mM dithiothreitol (DTT). Samples were then separated by size in porous 12% polyacrylamide gel (1D SDS polyacrylamide gel electrophoresis), under the effect of an electric field. The electrophoresis was run under constant voltage (130 V, 45 min) and the gels were stained using a Coomassie Blue solution (0.1% Coomassie Brilliant Blue R-250, 50% methanol, and 10% glacial acetic acid) and faded with the same solution lacking the dye. Densitometry was performed using ImageJ analysis software (NIH). This technique allows a quantitative digital analysis of image data from electrophoresis gels[29], calculating in this way the final antibody coverage degree of every immuno-NC.

**Physicochemical Characterization.** The hydrodynamic mean diameter of the NC was determined by PCS, using a 4700C light scattering device (Malvern Instruments, U.K.) and working with a He−Ne laser (10 mW). The light scattered by the samples was detected at 173°, and the temperature was set at 25 °C. The diffusion coefficient





measured by dynamic light scattering can be used to calculate the size of the $O^2$LNC by means of the Stokes−Einstein equation. The homogeneity of the size distribution is expressed as a polydispersity index (PDI), which was calculated from the analysis of the intensity autocorrelation function[36]. Electrophoretic mobility ($\mu_e$) as a function of pH was measured after diluting a small volume of the $O^2$LNC stock (with a total surface equal to 0.05 m2) in 1 mL of the desired buffered solution. It should be noted that all the buffers used in the $\mu_e$ studies had identical ionic strengths, being equal to 0.002 M except for measurements with PBS (pH 7.4 and 150 mM) and with RPMI cultured medium supplemented with 10% FBS (pH 7.4 and 150 mM). The $\mu_e$ measurements were made in triplicate using a nanozeta dynamic light-scattering analyzer (Zeta-Sizer Nano Z, Malvern Instruments, U.K.).

***In Vitro* Cellular Uptake.** Cells (1.5 × 105) from both BxPC-3 cells growing in monolayer and BxPC-3 PCSCs were seeded into Corning cell culture flasks (VWR, Spain) and into Ultra-Low Attachment 6-well plates (Corning, Spain), respectively. In this study, $O^2$LNCs were formulated by adding NR into olive oil at a concentration of 0.025% (w/w). Regarding Nile red, we assumed that because it is a lipophilic dye, it would be fully incorporated into the olive oil core of the $O^2$LNCs and would be very unlikely to be excluded. In any case, we performed an assay to evaluate the possible exclusion of Nile red from the $O^2$LNCs. Briefly, several aliquots of NR-$O^2$LNC were incubated at 37° for 4, 8, and 24 h in PB, PBS, and phenol red-free culture medium with FBS. These samples were centrifuged in Vivaspin tubes (1000 kDa pore size), and supernatants were collected and analyzed using a spectrofluorometer. Then, the same protocol described above was used to coat the $O^2$LNC surface by our antibody. The membrane binding and cell internalization of both labeled naked NR-$O^2$LNC and NR-αCD44-$O^2$LNC were examined by laser-scanning confocal microscopy. Briefly, the two





types of cell populations ($3 \times 10^3$/well) were seeded in Slide 8-Well chambers (ibiTreat, IBIDI) with RPMI medium and F12 spheres medium supplemented as described above, for the differentiated PCCs and the PCSCs, respectively. After 24 h, $4.4 \times 10^{11}$ of $NRO^2LNC$ and $NR-\alpha CD44-O^2LNC$ were added to the cells. The images were taken after 3 h of incubation. Imaging experiments were conducted with a Zeiss LSM 710 laser-scanning microscope using a tissue culture chamber (5% $CO_2$, 37 °C) with a plan-apochromat 63×/1.40 Oil DIC m27. Images were processed with Zen Lite 2012 software.

Additionally, to quantitatively analyze the uptake of the $O^2LNC$ in the two PC cell populations, a flow cytometry assay was performed. Once the different cell populations were obtained, these were washed and centrifuged at 1500 rpm for 5 min in tubes and resuspended in a 1% BSA (Bovine Serum Albumin) solution to block the possible non-specific binding of antibodies Next, $1x10^6$ cells per sample were incubated with $4.4x10^{11}$ $NR-O^2LNC$ and $NR-\alpha CD44-O^2LNC$. The incubation times of the cells with the $O^2LNC$were 15, 30 and 60 min. Then, the cells were washed with 1X PBS and centrifuged at 1500 x G for 5 min twice to remove the non-internalized NCs. Finally, the cells were resuspended in 300 µl of 1X PBS and analyzed for red and green fluorescence by flow cytometry (FACS CANTO II (BD Biosciences)) using the FACS DIVA software. Moreover, we performed a characterization of BxPC-3 PCCs and BxPC-3 PCSCs in order to corroborate their stemness properties. Briefly, cell surface marker levels of BxPC-3 PCSC were determined with human antibodies anti CD44-FITC and anti CXCR4-APC; and ALDEFLUOR assay (Stem Cell Technologies) to detect enzyme ALDH1 activity was performed to complete the characterization. Samples were measured and analyzed by flow cytometry on a FACS CANTO II (BD Biosciences). All experiments were performed in triplicate and replicated at least twice.





Sterility evaluations of all nanosystems were performed prior to develop $O^2LNC$ uptake studies in order to exclude possible biological contamination. Cells treated with naked $O^2LNC$ were used as controls.

Finally, in order to corroborate that the NR-$\alpha$CD44-$O^2LNC$ enter into the BxPC3 PCSCs (high CD44 expression) through a receptormediated mechanism (CD44 antigen), an antigen block assay by flow cytometry was performed. Once the cell subpopulation (BxPC3 PCSCs) was obtained, these were washed and centrifuged at 1500rpm for 5 min in tubes and resuspended in a 1% BSA solution to block the possible nonspecific binding of antibodies; next, $1 \times 106$ cells per sample were incubated with $4.4 \times 1011$ NR-O2LNC and NR-$\alpha$CD44-$O^2LNC$. The incubation times of the cells with the $O^2LNC$ were 30 min. Additionally, $1 \times 106$ cells per sample were incubated with 6 $\mu$L of anti-CD44 at 4 °C for 15 min. Afterward, the samples were incubated with $4.4 \times 1011$ NR-$O^2LNC$ and NR-$\alpha$CD44-$O^2LNC$. The incubation time of the cells with the $O^2LNC$ was 30 min. Then, the cells were washed with $1\times$ PBS and centrifuged at 1500g for 5 min twice to remove the noninternalized NCs. Finally, the cells were resuspended in 300 $\mu$L of $1\times$ PBS and analyzed for red and green fluorescence by flow cytometry (FACS CANTO II, BD Biosciences) using FACS DIVA software. All the experiments for PCCs were carried out in the presence of FBS except for PCSCs, which must be cultured without FBS.

**Encapsulation Efficiency, Drug loading and Retention Time of $O^2LNC$-PTX.** In order to know if our $O^2LNCs$ are a good drug vehicle nanosystem, two parameters were calculated, such as drug loading (DL) (1) and encapsulation efficiency (EE) (2). Then, a defined amount of PTX was dissolved in the olive oil and the synthesis of the $O^2LNCwas$ performed. Briefly, PTX was dissolved in olive oil at different concentrations and EE, as well as the permanence over time within the oil core of the





$O^2$LNC, was calculated. After the synthesis and the covalence protocol, $O^2$LNC, $O^2$LNC-PTX and αCD44-$O^2$LNC-PTX were stored at 4°C and aliquots were collected at different times (0, 1, 2, 3 and 4 weeks). These aliquots were centrifuged in Vivaspin 20® tubes with a pore size of 300 KDa at 2500xG for 15 minutes. Then, each aliquot was dissolved in 2-propanol, mixed by vortex and sonicated for 30 mins and centrifuged at 14000 rpm for 20 minutes. The supernatant was collected and analyzed by MS with a mobile phase of acetonitrile-water.

$$\%DL = \frac{Drug\ encapsulated(g)}{Total\ Nanocapsule\ weight(g)}\ x\ 100 \qquad (1)$$

$$\%EE = \frac{Drug\ encapsulated(g)}{Theoretical\ drug(g)}\ x\ 100 \qquad (2)$$

***In Vitro* Antitumor Performance.** The effect of PTX-loaded $O^2$LNC on cell viability was assessed using the MTT or CCK8 kit assay, depending on the cell population. In the case of differentiated BxPC-3PCCs, cells were seeded into culture flasks with 15 ml RPMI supplemented 10% of FBS. When confluence was optimal (80%), cells were detached with trypsin/EDTA and seeded into 96-well plates at a concentration of 3000 cells per well. The cells were then treated with $O^2$LNC-PTX, αCD44-$O^2$LNC-PTX and free PTX for three days. Free PTX was dissolved in DMSO, and PTX-loaded $O^2$LNCs were diluted in RPMI medium. The cytotoxicity evaluation was performed by MTT assay according to the protocol. Briefly, the MTT solution was prepared at 5 mg ml$^{-1}$ in 1X PBS and then diluted to 0.5 mg/ml in MEM without phenol red. The sample solution in the wells was removed and 100 µl of MTT dye was added to each well. Plates were shaken and incubated for 3 h at 37◦C. The supernatant was removed and 100 µl of pure DMSO was added. The plates were gently shaken to solubilize the formazan that was





formed. The absorbance was measured using a plate reader at a wavelength of 570 nm. In the case of BxPC-3 PCSCs, cells were seeded into Ultra-Low Attachment 96-well plates at a concentration of 5000 cells per well. After 3 days, cells were treated with free PTX, $O^2$LNC-PTX and $\alpha$CD44-$O^2$LNC-PTX and the culture medium used was F12 supplemented as described above. The toxicity evaluation was performed by CCK8 assay according to the protocol. Briefly, 15 µl of the CCK 8 kit was added to each well. Plates were shaken and incubated for 3 h at 37◦C. The absorbance was measured using a plate reader at a wavelength of 492 nm. The inhibitory concentration 50 (IC$_{50}$) values were calculated from dose-response curves by linear interpolation. All of the experiments, plated in triplicate wells, were carried out at least three times, and for PCCs, cultures were done in the presence of FBS and PCSCs without FBS.

**Dual GFP-NanoLuc LVs Production.** The SELWP is a self-inactivated (SIN) LV expressing GFP-2A-NanoLuc under the control of an internal spleen focus-forming virus (SFFV) promoter. SELWP was generated previously (Tristán-Manzano et al, unpublished) based on the SEWP LV[37]. LV particles were generated by transient co-transfection of 293T cells using eGFP-NanoLuc vector plasmid together with the packaging plasmid pCMVDR8.91 (http://www.addgene.org/Didier_Trono) and the p-MD-G plasmid encoding the vesicular stomatitis virus (VSV-G) envelope gene (http://www.addgene.org/Didier_Trono). Vector production was performed as previously described[38]. Briefly, 293 T cells were planted on petri-dishes (Sarsted, Newton, NC), in order to unsure 80% of confluence for transfection. The vectors, the packaging and envelope plasmid (in proportion 3-2-1) were resuspended in 45 µl of LipoD293 (SignaGen, Gainthersburg, MD, USA). The plasmid-LipoD293 mixture was added to pre-washed cells and incubated for 6–8 h. After 48 hours viral supernatants were collected, filtered through 0.45 µm filter (Nalgen, Rochester, NY), aliquoted, and





storage at -80ºC. Viral titers were determined by transducing 293T cells with different volumes of supernatant. The percentage of eGFP$^+$ cells was determined by flow cytometry 72h later and transducing units per ml (TU/ml) were estimated according to the formula: $[(10^5$ plated cells $\times\%GFP^+$ cells$) \times 1000]/\mu l$ of LVs.

**BxPC-3Cells Transduction.** BxPC-3 PCCs (100,000) were seeded in a 24-well plate and incubated with SELWP LV supernatant at a Multiplicity of Infection (MOI) of 3 during 5h at 37ºC and 5% $CO_2$. eGFP expression was determined by flow cytometry (FACS Canto II, Becton Dickinson, New Jersey, US) at 72h after transduction. Additionally, cells were stained with anti-CD44-APC and CXCR4-PE for 20 min at room temperature.

***In vivo* Assay for Targeted PCSCs.** All *in vivo* experiments were performed in male NOD scid gamma mice (NSG, NOD. Cg-Prkdcscid Il2rgtm1Wjl/SzJ). Animal welfare and experimental procedures were carried out in accordance with institutional (Research Ethics Committee of the University of Granada, Spain) and international standards (European Communities Council directive 86/609). All animals (n= 5 per group) were maintained in a micro-ventilated cage system with a 12h light-dark cycle, and they were manipulated in a laminar air-flow cabinet to keep to the specific pathogen-free conditions. To establish orthotopic xenograft tumors, six- to eight-week old male mice were used. Mice were anesthetized by intraperitoneal administration of ketamine (4 mg/kg) and midazolam (80 mg/kg), we performed a medial laparotomy from the xiphoid appendix to the lower left third of the abdomen. The retrogastric space was accessed and the tumors were generated into the tail of the pancreas by injections of $2\times10^5$ BxPC-3-SELWP PCSCs/mouse mixed with Corning ®Matrigel® Matrix and using 26-gauge needles. Then, the mice were sewn with 5/0 thread and administered analgesics and antibiotics. Once the generated tumor had a size of around 0.5 cm$^3$,





intravenous administration in the vein tail of the mice of 100µl of PBS and 100µl of $O^2$LNC-i780 and αCD44-$O^2$LNC-i780 at a concentration of 1,1 x $10^{13}$ $O^2$LNC/ml was performed in a single dose and then the mice were evaluated by IVIS® (Perkin Elmer). The mice were left for a week in order to eliminate the $O^2$LNCs that were not specifically bound to the tumor tissue. During that week, the bioluminescence with a previous intraperitoneal administration of NanoGlo substrate (Promega) in PBS and the fluorescence of the mice was analyzed in the IVIS® on three alternate days. Finally, mice were euthanized by cervical dislocation and tumors and organs were excised, photographed, and analyzed by IVIS.

**Statistical analysis**

The data are presented as the mean ± the standard deviation in the error bars. The sample size (n) indicates the experimental repeats of a single representative experiment, with 3 being unless otherwise specified. The results of the experiments were validated by independent repetitions. Graphs and statistical difference data were made with GraphPad Prism 6.0 (Graphpad software Inc.). Statistical significance was determined using Student's t-test in paired groups of samples with known median A p-value of ≤ 0.01 was considered significant.

**Results and Discussion**

**Preparation and Physico-Chemical Characterization of Nanocapsules and Immuno-Nanocapsules.**

*Preparation of NCs*. These ($O^2$LNC) were prepared by using a modified protocol based on the solvent-displacement technique reported by Calvo et al[35]. Briefly, it consists of





the formation of a nanoemulsion by incorporating, with a high mechanical force, an organic phase into an aqueous phase under stirring. This produces a nanoemulsion of a certain size due to the addition of a defined amount of olive oil to the organic phase (Figure 1A).

*Particle Size and Stability Over Time.* The synthesis method produced a homogeneous population of $O^2LNC$ with a concentration of 1,1 x $10^{13}$ $O^2LNC$/ml and with an average diameter of around $111 \pm 18$ nm and a polydispersity index (PDI) of around $0.10 \pm 0.02$. This size is lower than those previously reported using a similar technique and similar shell components[29]. The size reduction is related to the increase of the mechanical energy applied when organic and aqueous phases are mixed. The use of high-speed homogenization to produce small particle sizes has already been reported[39]. Additionally, it is important to remark that the size reduction does not affect the colloidal stability of the sample and the $O^2LNC$ size remained at a constant value under storage conditions, i.e. pure water or phosphate buffer (PB) and 4 °C for at least 4 months. Each week for at least 4 months, the size and PDI of the same $O^2LNC$ samples were analyzed, obtaining a similar size to those taken after synthesis (data not shown). For in vivo applications, the specialized literature advises a mean diameter around 200 nm being a nanosystem with sizes between 20 and 100 nm that show the higher potential[40]. Therefore, the $O^2LNC$ diameter is optimal for in vivo applications, where the size and size distribution are a decisive variable in ensuring adequate biodistribution by crossing biological barriers and minimizing the macrophage uptake[41].

*Chemical Characterization of $O^2LNC$ by Nuclear Magnetic Resonance Spectroscopy.* Various Nuclear Magnetic Resonance (NMR) spectroscopy experiments were performed in order to verify that all the components used in the synthesis of the $O^2LNC$ were integrated in the nanocapsule entities comprising the colloidal nanoemulsion





solution. Before analysis, $O^2$LNC was further cleaned by centrifugation in Vivaspin 20®

tubes with a pore size membrane of 300 KDa to ensure an essentially complete

elimination of the organic solvents (ethanol and acetone) employed in their synthesis.

The $^1$H NMR spectrum of intact $O^2$LNC (Figures S1 and S2), diluted to a final 13 %

content of $D_2O$, is clearly dominated by the signals of the olive oil core of the $O^2$LNC,

as can clearly be seen by direct comparison with the NMR spectrum of olive oil in

$CDCl_3$ (Figure S3). The observation of olive oil signals NMR in an aqueous sample

provides, by itself, direct evidence of the emulsion nature of the sample since olive oil is

completely immiscible with water. On the other hand, the linewidth of the observed

olive oil signals in the $O^2$LNC sample is much broader than the linewidth of olive oil

signals in $CDCl_3$. This feature is partially the consequence of the higher viscosity of the

nanoemulsion but, more importantly, is the result of a reduced transverse relaxation

time ($T_2$) product of a much slower tumbling rate of the triglyceride molecules of olive

oil, when confined in the nanoparticle core, compared to the tumbling of the same

molecules when freely dissolved in $CDCl_3$. Additionally, and as final evidence,

qualitative diffusion NMR experiments proved that the olive oil signals do not diffuse at

the expected rate for free triglyceride molecules in solution but rather at a very slow

rate, a feature compatible with the size of the NCs[42] (Figure S4). Once again this is a

consequence of the confinement of olive oil in the core of the $O^2$LNC. Regarding the

other chemical constituents of the $O^2$LNC, only the signals from the surfactant Pluronic

F-68 could be observed in the $^1$H NMR spectrum of intact $O^2$LNC (Figure S4), while no

signals related to Epikuron or DCA are detected in the spectrum. These three

components (surfactant, phospholipid and deoxycholic acid) comprise the shell of the

$O^2$LNC and as such conform to a rather rigid supramolecular structure expected to

display a very short $T_2$ broadening the corresponding NMR signals beyond detection.





The fact that the signals of the oligomeric surfactant Pluronic F-68 can still be observed is a consequence of the inherent flexibility of the part of these molecules not rigidly embedded in the $O^2LNC$ shell but rather exposed to the bulk water.

To further demonstrate the chemical nature of the components integrated in the $O^2LNC$, these were disintegrated by dilution with isopropanol and sonication. After the disintegration of the $O^2LNC$, the sample was split into two aliquots which were concentrated to dryness and resuspended, one in $CDCl_3$ and the other in $CD_3OD$ for $^1H$ NMR analyses. Comparison with the corresponding reference spectra of olive oil, Pluronic F-68, Epikuron and deoxycholic acid acquired in these two solvents unambiguously demonstrated the presence of these components in the original $O^2LNC$ sample (Figures S5-S15).

*Synthesis of the Olive Oil Immuno-NCs.* In order to produce the immuno-NCs, a defined amount of anti-CD44-FITC (αCD44) antibody equal to 0.18 mg/m$^2$ was initially incubated with the $O^2LNC$. This amount of antibody, corresponding to a low coverage degree, was used due to the limited protein concentration commercially available and because the density of the molecules on the surface of the NPs affects their affinity for the substrate. Thermodynamically, the binding of a ligand to its substrate facilitates the subsequent binding of its neighboring ligands[43]. Biologically, the multiple interactions of the NP with the cell membrane forces the clustering and local concentration of receptors. This triggers the membrane envelope and leads to internalization[44]. This allows the use of multiple ligands of relatively low affinity to bind their target efficiently and with great appetite[45]. However, this increase in affinity is not always linear. In some cases, the cooperative effect of the ligand can saturate and further increase the density of the ligand causing harmful effects on cell binding[46,47]. This effect can be explained by an incorrect orientation of the ligand, steric impediments caused by







neighboring molecules or competitive behaviors with other molecules for binding with the receptor. The covalent coupling was developed to identify the optimal experimental conditions for a ECDI procedure adapted to the properties of CD44 antibody, as is described in detail in the Experimental Section. No aggregation of the nanosystem was observed during the experimental period, in spite of the well-established idea about the immobilization of antibody molecules on the surface of nanoparticles that strongly alters their surface properties, with a possible modification of the colloidal stability of the nanosystem influenced by the coverage degree and the suspension medium. Although the surface antibody layers drive a change in the physicochemical properties of $O^2$LCN's original surface, the low coverage degree, which preserves the electrostatic repulsion between free ionized carboxilic surface groups (see the electrokinetic characterization), and the optimization of the medium conditions, pH, and ionic strength throughout the coupling process prevent $O^2$LCN aggregation[29,48]. Figure 1A shows a schematic representation of the coupling process to produce the immuno-$O^2$LNC after incubation time, the αCD44-$O^2$LNC complex was separated from the unbound protein by a dialysis procedure, which is normally used when working with nanoemulsions and soft particle systems[49], and has previously shown their efficacy working with the same type of NCs[33]. The size of the formulation (αCD44-$O^2$LNC) with the initial theoretical antibody coverage (low) at pH 7.4 and low-ionic strength medium remained similar to those of the original $O^2$LNC systems. A protein quantification analysis by means of spectrophotometric assays of the first dialysis elution volumetric fraction did not detect any presence of antibody molecules, which could mean a total surface immobilization of the initial antibody amount. The presence of IgG on the αCD44-$O^2$LNC surface was estimated by SDS gel chromatography under reducing conditions that confirmed the immobilization of the antibody on the $O^2$LNC surface (Figure 1B). Gel electrophoresis





shows a migration response typical for IgG antibodies, with two heavy and two light chains corresponding to an upper band at ~50 kDa (for the Fc fragment) as well as a lower band at ~25 kDa (for the Fab moieties).50 As shown in Figure 1B, the intensity of the band corresponding to αCD44-O$^2$LNC (lane 2) was similar to that of a free αCD44 antibody (lane 1). Additionally, a quantification of the surface amount of the antibody was carried out by means of a densitometric procedure, and the coupling efficiency reached a value of 93.4% for the initial theoretical coverage of 0.18 mg/m$^2$. In previous studies carried out in our research group with a similar type of NCs, the coupling efficiency reached values of around 50−70% for an initial antibody theoretical coverage of 2.5 mg/m$^2$ [29]. Moreover, Goldstein et al. showed a low coupling efficiency when working with monoclonal antibodies and Fab' fragment on functionalized NCs as a consequence of the presence of polyethylene glycol chains on the surface[49]. It is amply documented how the presence of PEO layers is frequently employed to reduce protein binding[51,52], and the effect on the protein adsorption processes is clearly reflected by a reduction in the amount of adsorbed proteins mediated by the presence of surface poloxamer molecules[53]. Nonetheless, we have optimized the methodology and have achieved a coupling near 100% of the antibody molecules to the surface of the O$^2$LNC using a very low protein coverage (0.18 mg/m$^2$) rendering a stable nanoemulsion.

It was previously contrasted, working with four different specific antibodies, how this covalent coupling method using ECDI and carboxylic surface groups allows a directed covalent bound of antibody molecules to the surface, preserving an adequate immunological activity over the corresponding antigen molecules. However, the same coupling procedure in the absence of ECDI, in which antibodies result in being physical adsorbed on the LNC surface, reduce drastically the later specific immunological.29





Furthermore, the presence of surface poloxamer molecules, the low initial coverage degree, and the low affinity of αCD44 antibody molecules for the negatively charged surface in the experimental conditions[54] suggest that antibody adsorption is very residual in our immune nanosystem.

**Table 1.** *Size and PDI of O2LNC and αCD44-O2LNC Incubated with Different Mediums.*

| | $O^2LNC$-PB | αCD44-$O^2LNC$-PB | $O^2LNC$-PBS | αCD44-$O^2LNC$-PBS | $O^2LNC$-FBS | αCD44-$O^2LNC$-FBS |
|---|---|---|---|---|---|---|
| **Size (nm)** | 110±20 | 110±20 | 110±20 | 110±20 | 120±10 | 110±20 |
| **PDI** | 0.10±0.02 | 0.09±0.01 | 0.15±0.01 | 0.16±0.01 | 0.19±0.04 | 0.25±0.01 |





**A**

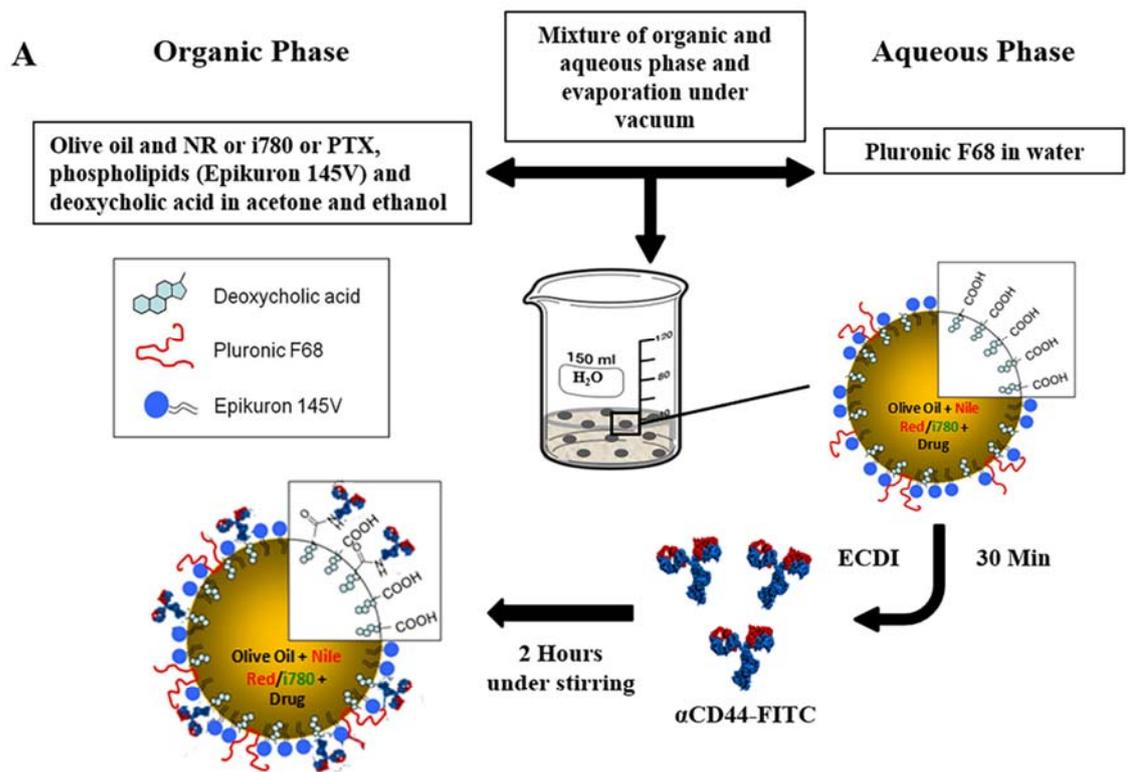

**B**

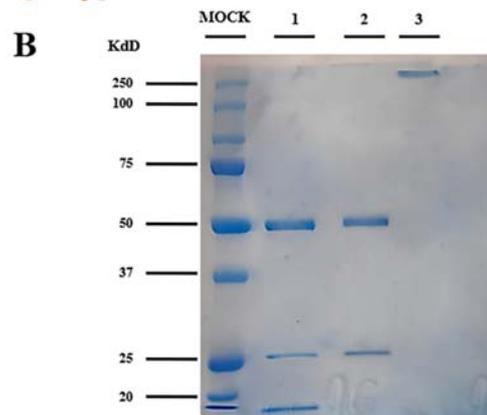

**C**

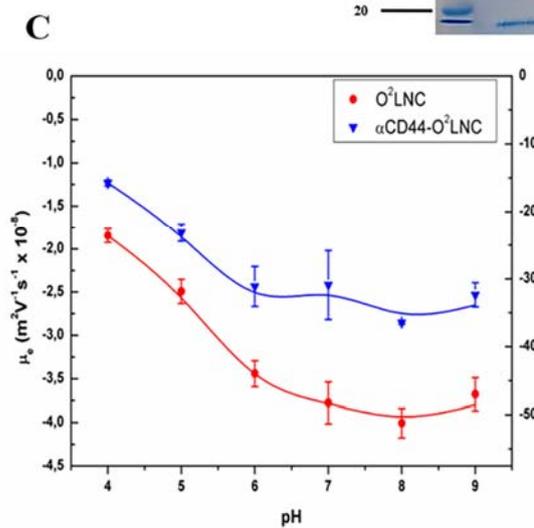

**D**

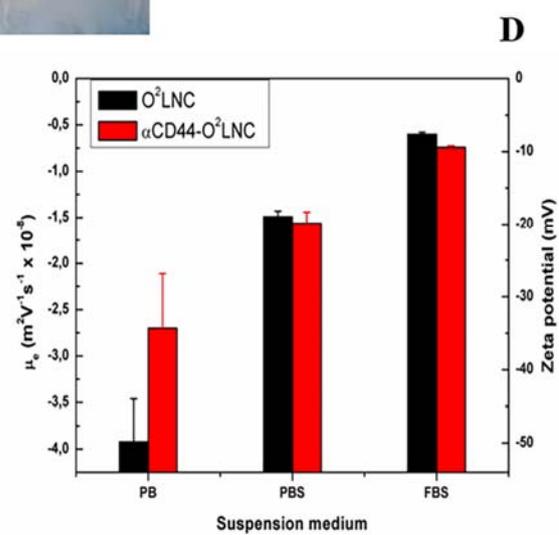





**Figure 1.** *Synthesis and physicochemical characterization of $O^2$LNC and αCD44-$O^2$LNC. (A) Scheme of synthesis and functionalization of $O^2$LNC and αCD44-$O^2$LNC. (B) SDS-PAGE of αCD44 (lane 1), αCD44-$O^2$LNC (lane 2), and bare $O^2$LNC (lane 3). (C) Electrophoretic mobility and Z potential of O2LNC and αCD44-$O^2$LNC in a pH range from 4 to 9. (D) Electrophoretic mobility and Z potential of $O^2$LNC and αCD44-$O^2$LNC incubated in PB at pH 7.4 (PB), PBS, and RPMI culture medium supplemented with 10% FBS.*

**Electrokinetic characterization and colloidal stability.**

Usually, when protein molecules coat the surface of colloidal particles, their electrokinetic behavior change markedly compared to the same bare surfaces [54–56]. The most representative electrokinetic parameter, the zeta potential, was determined in low-ionic strength media at different pH values from 4.0 to 9.0, through the measurement of the electrophoretic mobility, $\mu_e$ (Figure 1C). In terms of the electrokinetic behavior of bare $O^2$LNC, the $\mu_e$ results were in agreement with the nature of the shell of these NCs in which carboxylic groups predominate on the surface, showing the typical behavior of colloids with weak acid groups, that is, constant $\mu_e$ values at basic and neutral pH values that begin to fall to acidic pH around the pKa (4.8).[48] Nevertheless, the most important aspects of the $\mu_e$ results are related to the presence of IgG molecules immobilized onto the αCD44-$O^2$LNC surface, which significantly alter the mobility data and indicate a clear correspondence with the adhered antibody and its IEP[54] (Figure 1C). Additionally, it is well known that when a protein covers colloidal particles, the IEP of such complexes also depends on the degree of protein coverage, so that it gradually tends to the pure protein IEP when the protein load on the nanoparticles increases[49,55]. The Z potential of the $O^2$LNC and αCD44-$O^2$LNC measured at pH 7.4 was around −50 and −35 mV, respectively (Figure 1C). This fact demonstrates that the





covalent binding of the αCD44 antibody was carried out satisfactorily and positive antibody molecules at acid pH partially screen the negative surface charge of the O2LNC, reducing the absolute Z potential values under the IEP of the protein.

On the other hand, the colloidal stability of delivery nanosystems in physiological media is crucial to successfully achieve their biological applications. As can be observed in Figure 1C, $O^2LNC$ and αCD44-$O^2LNC$ have electrokinetic behavior in mediums with low ionic strength, from neutral to basic pH, without variations in the negative surface charge, which can be correlated to stable nanosystems because of an electrostatic repulsion mechanism[56]. Thus, the size for both $O^2LNC$ and αCD44-$O^2LNC$ nanosystems remained constant throughout this pH range. Additionally, the colloidal stability was checked in mediums for physiological conditions. Therefore, electrokinetic parameters, hydrodynamic size, and PDI were measured for $O^2LNC$ and αCD44-$O^2LNC$ by incubating them in physiological medium (PBS) and in cell culture medium (FBS) (RPMI supplemented with 10% FBS). When both nanosystems were in contact with PBS and FBS, the Z potential decreased their negative absolute values (Figure 1D) as a consequence of the salinity of the medium (150 mM), which could mean the aggregation of the NCs. However, as can be seen in Table 1 and Figure S16, neither NCs did change their sizes and both remained stable in these media. Furthermore, both NC systems also keep a narrow size distribution (PDI ≤ 0.2). A stabilization mechanism mediated by hydration forces could be responsible for these results. This stabilization phenomenon is typical for hydrophilic colloidal systems at high salt concentrations[57]. In our nanosystems, PEO chains from poloxamers and antibody molecules supply the hydrophilic character to the surface shell and this behavior has been previously revealed by NCs with a similar shellcomposition[58,59] in complex mediums with different electrolytes.





**Drug Loading and Encapsulation Efficiency**.

DL and EE were calculated to know whether O2LNCs are good drug delivery vehicle nanosystems. In order to quantify these parameters, a defined amount of PTX was dissolved in the olive oil and the synthesis of the $O^2$LNC-PTX and αCD44-$O^2$LNC-PTX was performed. Then, the real amount of PTX inside the NCs measured by MS showed that our NCs had 2.2% of DL and 81.1% of EE. In addition, a 4-week test was conducted to observe the amount of PTX remaining within our O2LNCs before and after they undergo the covalence process with the αCD44 antibody (see details in the Experimental Section). In this way, at time 0, the real amount of PTX within the O2LNC oily core was around 80%, and after the covalence process, this amount was reduced to around 40% (Figure S18A). This decrease is because of the loss of PTX during the stirring of the covalence protocol and in the number of O2LNCs during the dialysis process by centrifugation in Vivaspin 20 tubes. It should be noted that as the weeks progress, the amount of PTX in both types of O2LNC remains practically constant, with a difference of 10% between week 0 and week 4 (Figure S18A).

***In vitro* Cellular Uptake of Immuno-NCs.**

The cellular uptake of both $O^2$LNC and αCD44-$O^2$LNC was investigated on BxPC-3 PCC growing in monolayer (low expression of CD44) and on BxPC-3 PCSC secondary spheres (high expression of CD44) by both confocal microscopy and flow cytometry. It is well known that the levels of CD44 expression in BxPC3 PCSCs are much higher than in their tumor counterpart[60-62]. For this purpose, $O^2$LNCs were labeled with Nile Red (NR) to investigate the cellular entry of our nanosystems. No fluorescence was detected in any of the samples measured by spectrofluorometry (data not shown). In addition, our research group published an article where they evaluated the





internalization of NR and its fluorescence intensity in different types of nanoparticles, including the ones shown here[33]. First, we performed a characterization of the expression levels of CD44 and CXCR4 markers in our tumor cell line ((monolayer and PCSCs) (Figure2A)) as well as the determination of the ALDH1 enzymatic activity. The overexpression of these markers, as well as the increase in ALDH1 enzymatic activity, is an indication that tumor cells have stem properties[34]. The expression percentages obtained were 6.6%, 16.7% and 68.7% in monolayer and 50.3%, 97.8% and 86.4% in secondary spheres for ALDH1, CD44 and CXCR4 respectively (Figures 2B,C). Control DOTS PLOTS are shown in Figure S17. These percentages were satisfactory and can thus conclude that a population of BxPC-3 PCSCs obtained was valid for the study. Confocal microscopy images show how NR-O$^2$LNC enters into both types of cell populations after 3 hours of accumulating NR in the cytoplasm (Figure 2D). In the case of NR-αCD44-O$^2$LNC, where CD44 was labeled with FITC, we observed the specific recognition of the CD44 membrane receptors by these NCs and the release into the cells of NR after 3 h of incubation (Figure 2D). We incubated the cells with our O$^2$LNC-based nanosystems for 3 h because it was the optimal time to visualize the fluorescence by confocal microscopy. This optimal time was established based on our previous experiments where NR-O$^2$LNC penetrated into monolayer cells and secondary spheres instantaneously, obtaining optimal fluorescence images (data nor shown). On the contrary, NR-αCD44-O$^2$LNC penetrated into monolayer cells after 3 h of incubation in order to obtain optimal fluorescence images. These NR-αCD44-O$^2$LNC penetrated the secondary spheres after 15 min of incubation, which suggest a specific recognition of the CD44 receptors for NR-αCD44-O$^2$LNC. These results may suggest that our nanosystem enters into the cells by a clathrin-mediated endocytosis mechanism, which has been reported as the main mechanism occurring with nanoparticles smaller than 200





nm.63 The flow cytometry assay confirmed the uptake and the specific recognition of our $O^2$LNC-based nanosystems in the two BxPC-3 PC cell populations. The analysis showed the differential uptake efficiency and specificity for NR-$O^2$LNC in monolayer BxPC-3 PCC and BxPC-3 PCSC subpopulation depending on whether they were functionalized or not with the αCD44 antibody. Moreover, because there were differences in the expression levels of the CD44 membrane receptor for both PC cell populations, we also found a different behavior for each type of the nanosystem. Thus, only around 17% of PCCs growing in the monolayer were positive for CD44, and almost all the PCSCs (about 98%) overexpressed this specific surface antigen (Figure 2C). However, most bare NR-$O^2$LNCs entered into both monolayer PCCs and PCSCs regardless of their CD44 expression level (Figure 2E,F). In contrast, for NR- αCD44-$O^2$LNCs, the entry was dependent on the CD44 expression in the membrane of the PC cells, with uptake values in PCSCs around 80, 86, and 85% at 15, 30, and 60 min, respectively, and only around 5% in PCCs growing in the monolayer (Figure 2E,F). This suggests that our NR-αCD44-$O^2$LNC actively recognizes the CD44 receptor overexpressed on PCSCs, producing the cell internalization of the nanosystem, contrary to what happens with PCCs (low CD44 expression), where these LNC displayed very low internalization. These results are in concordance with previous studies showing the targeted killing of PCSCs overexpressing the CD44 surface marker by NCs functionalized with hyaluronic acid (HA), the most common and immediate ligand for CD44.64,65 Finally, and in order to ensure that NR-αCD44-$O^2$LNC have selective targeting against PCSCs (high expression of CD44), the blocking assay showed a 50% reduction in the percentage of cells positive for NR-αCD44-$O^2$LNC (Figure 2G). Therefore, we can confirm that our αCD44-$O^2$LNCs were able to selectively bind to CD44 receptors releasing their content inside the PCSCs[66]. It is worth highlighting the





selective uptake efficiency of NR-αCD44-O$^2$LNC in the BxPC3 PCSCs as opposed to the more differentiated tumor and attached cell population. Recently, Trabulo et al. published a study where they covalently attached an anti-CD[47] antibody to the surface of iron oxide nanoparticles for the recognition of this membrane receptor overexpressed in PCSCs, obtaining very similar results to those described in this study[67]. Moreover, because the interaction of HA and CD44 promotes EGFR-mediated pathways, consequently leading to tumor cell growth, tumor cell migration, and chemotherapy esistance in solid cancers[23], the ability of our O2LNC functionalized with the αCD44 antibody could bind and neutralize the receptor by competitive inhibition of its HA ligand and consequently prevent the receptor-signaling cascade activation. All this makes our nanosystem, a good candidate for targeted therapy against PCSCs.





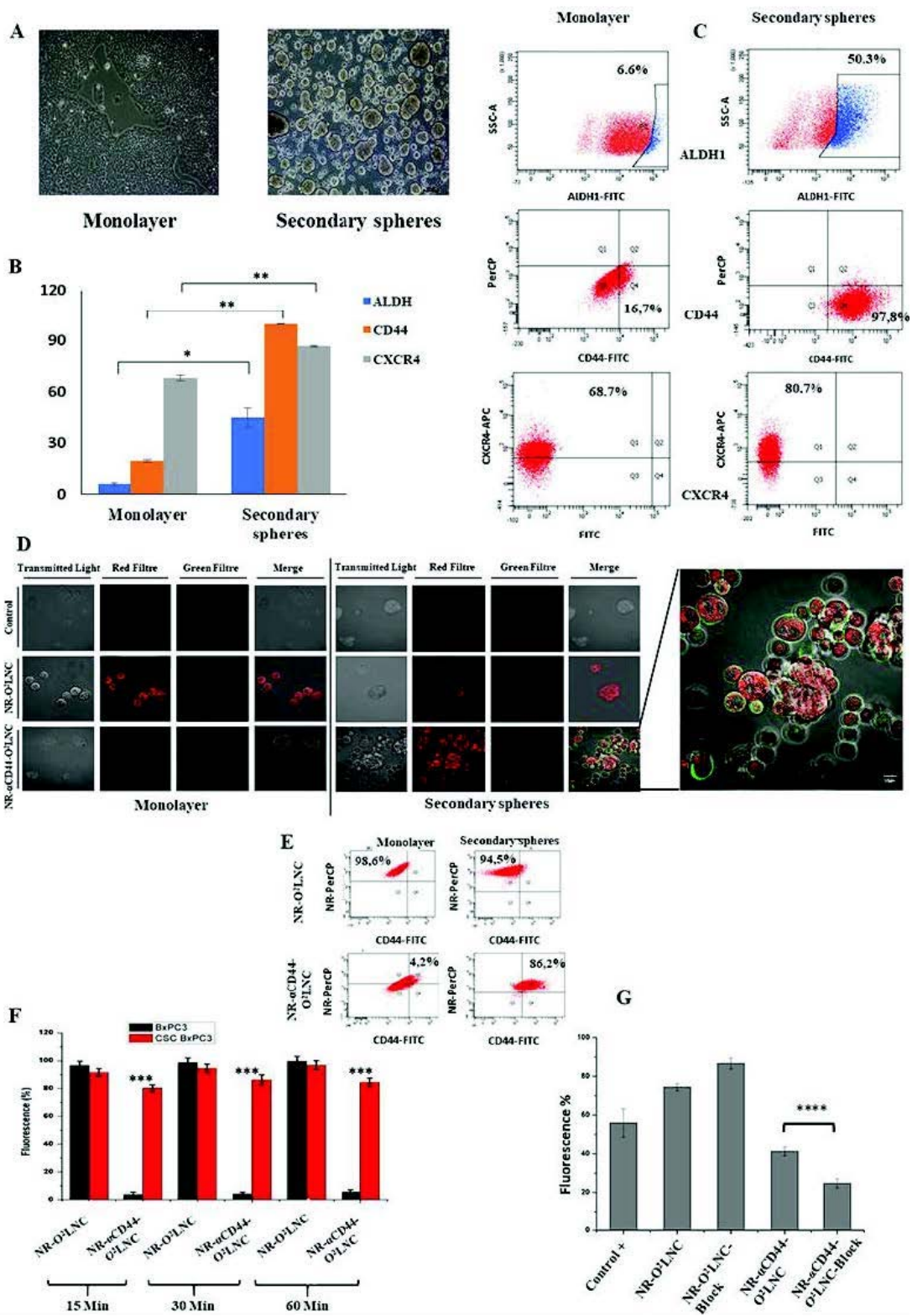





**Figure 2.** *Characterization and in vitro uptake of O²LNC and αCD44-O²LNC in both BxPC-3 PCCs and PCSCs. (A) Representative images of BxPC-3 PCCs (monolayer) and BxPC-3 PCSCs (secondary spheres). (B) ALDH1 activity, CD44, and CXCR4 surface expression in the monolayer and secondary spheres cell cultures. Data are graphed as mean ± SD and the significant values are calculated using the t-test (\*p < 0.01; \*\*p < 0.001). (C) Representative DOT PLOTS of BxPC-3 PCCs and PCSCs analyzed by flow cytometry. (D) Confocal microscopy of BxPC-3 PCCs (monolayer) and BxPC-3 PCSCs (secondary spheres) respectively (scale bar = 10 μm) incubated 3 h with NR-O²LNC and NR-αCD44- O LNC, respectively. Bare O2LNC were used as control. (E) Uptake of NR-O²LNC and NR-αCD44-O²LNC in BxPC-3 PCCs and BxPC-3 PCSCs analyzed by flow cytometry. Data are mean values ± SD. \*\*\*p < 0.001 shows the significant values calculated using t-test. (F) Representative DOT PLOTS of NR-O²LNC and NR-αCD44-O²LNC uptake in BxPC-3 PCCs and BxPC-3 PCSCs at 30 min analyzed by flow cytometry. (G) Blocking assay by uptake of NR-O²LNC and NR-αCD44-O²LNC in BxPC-3 PCSCs analyzed by flow cytometry. Data are mean values ± SD. \*\*\*\*p < 0.001 shows the significant values calculated using t-test.*

### *In vitro* antitumor cytotoxicity of paclitaxel-loaded O²LNC and αCD44-O²LNC

A cell viability assay was assessed to check the improvement of the antitumor activity of PTX by our functionalized O²LNC against differentiated PCCs and PCSCs. This test was performed comparing the effects of free PTX, O²LNC-PTX, or αCD44-O²LNC-PTX. The results showed that free PTX displayed significant differential antitumor potency depending on the PC cell population (Figure S18B). Thus, in BxPC-3 adherent PCC, the IC50 of free PTX was 1.7 nM; however, in BxPC-3 PCSCs, the IC50 value





was 133.4 nM (Figure S18B). This difference is due to the fact that PTX has a very high efficacy on differentiated and replicative tumor cells,68 but this efficiency drastically decreases in CSCs. Moreover, we observed that in monolayer BxPC-3 PCCs, neither O$^2$LNC-PTX nor αCD44-O$^2$LNC-PTX improved their antitumor effect but instead increased their IC50 slightly (Figure S18B). In contrast, although the O$^2$LNCPTX in the population of PCSCs very slightly reduced the IC50 value to 113.8 nM; however, αCD44-O2LNC-PTXs were able to reduce the IC50 down to 34.2 nM (Figure S18B). The increased potency of up to 4 times of αCD44-O$^2$LNC-PTX on PCSCs in comparison to free PTX, and up to 3.3 times with respect to O$^2$LNC-PTX, indicate the targeted activity against PCSCs overexpressing CD44, which facilitates the internalizing, favoring the bioavailability and cytotoxic effect of PTX. The significant reduction in cell viability provoked by αCD44-O$^2$LNC-PTX compared to naked O$^2$LNC-PTX is in agreement with other studies conducted with nanoparticles of a different nature, which are modified to be specifically targeted against PC cells[69–71]. Additionally, as shown in Figure S18C, we can see the cytotoxic effect of free PTX, O$^2$LNC-PTX, and αCD44-O$^2$LNC-PTX in the monolayer BxPC-3 PCCs and in the secondary sphere BxPC-3 PCSCs.





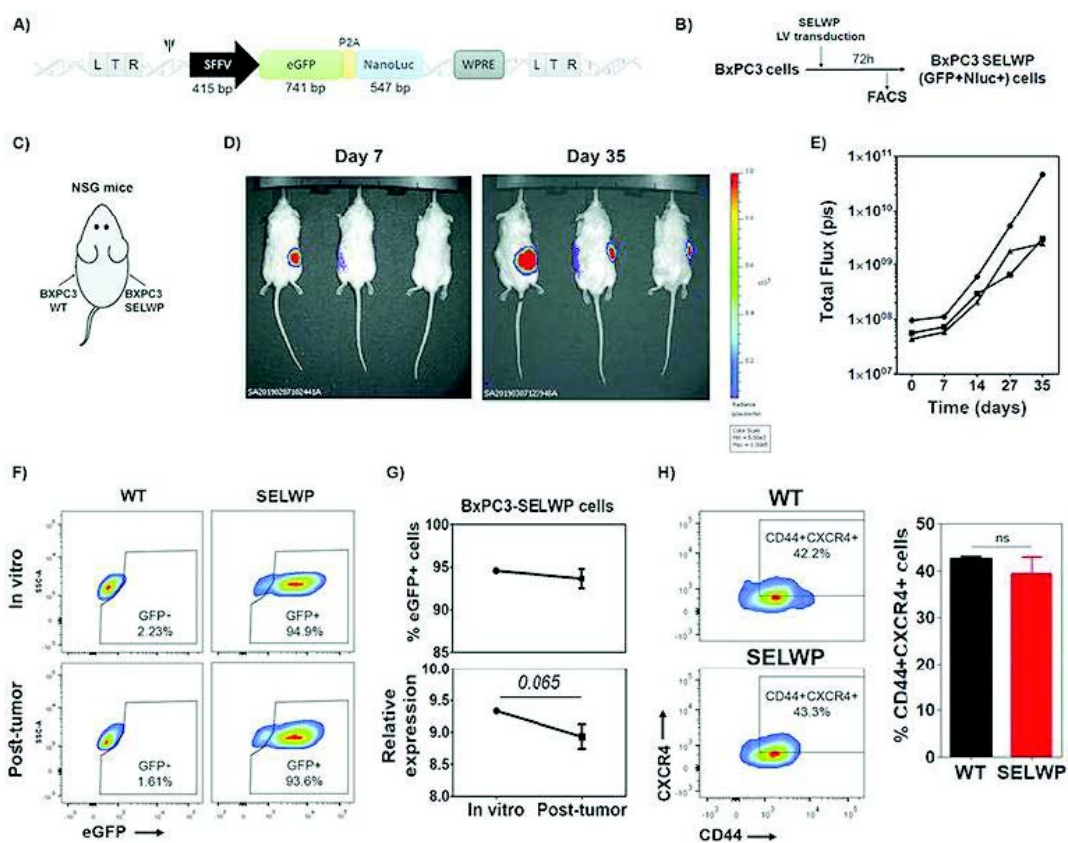

**Figure 3.** *BxPC-3 cells efficiently transduced with SELWP constitute a valid model for monitoring tumor growth in vitro and in vivo. (A) Representation of the SELWP LV encoding eGFP and NanoLuc under a SFFV promoter in a SIN LV backbone. (B) Scheme showing the procedure to generate BxPC-3-eGFP+/NanoLuc+ cells (BxPC-3-SELWP). (C) Scheme indicating the injection site of BxPC-3-WT and BxPC-3-SELWP to monitor tumor growth. Cells (300,000) in PBS were mixed with Matrigel (proportion 1:1) and inoculated in the left (BxPC-3-WT) and right (BxPC-3-SELWP) flank of 16-week old NSG mice. (D) Representative images of IVIS bioluminescence acquired at days 7 and 35 postinoculation after the administration of the NanoGlo substrate intraperitoneally. (E) Total photon flux analysis along time. (F) Representative plots showing eGFP expression in BxPC-3-WT (left plots) and BxPC-3-SELWP cells (right plots), before (top plots) and after (bottom plots) transplantation into mice for tumor*





*generation. (G) Graph showing percentage (top) and relative expression levels (bottom) of eGFP+ before (in vitro) and after (post-tumor) transplantation into mice. Relative expression was calculated as the median of FITC intensity of transduced eGFP+ cells/median of FITC intensity of nontransduced population. No significant p value, paired T-test, and two tails. (H) Representative plots (right) and graph showing the percentage of CD44+ CXCR4+ BxPC-3 cells isolated from WT and SELWP tumors (N = 3). CD44 and CXCR4 expression was determined by flow cytometry 7 days after sacrifice. No significant p value, paired T-test, two tails.*

## Luminescence Stable modification of BxPC-3 cancer cells for *in vivo* monitoring of αCD44-O²LNC-targeted efficacy

There is a need to create clinically relevant models for studying treatments directed against PC in established human PC orthotopic xenografts by the visualization of tumor growth in early stages and tracking tumor behavior and progression. Previous results indicate that the model is advantageous and several PCCs, including BxPC-3 PCCs, have been modified to express luciferase[72]. Thus, our aim was to generate a stable cell model that would allow monitoring pancreatic tumor growth in vivo and the characterization of potential phenotypic changes on the tumor cells. In addition, we decided to test the behavior of a novel small luciferase, small and highly penetrant NanoLuc[73,74], in combination with the expression of eGFP in a dual vector. We produced LV particles using the SELWPLV (Tristán-Manzano et al., unpublished) that allow both in vitro and in vivo monitoring of transduced cells by flow cytometry (eGFP) and bioluminescence imaging (NanoLuc), respectively (Figure 3A). BxPC-3 PCCs were transduced with SELWP LVs at a MOI of 3 to generate BxPC-3-SELWP





cell lines expressing both eGFP and NanoLuc. Inoculation of BxPC-3-SELWP in the left flank of immune-deficient NSG mice allowed the in vivo monitoring of tumor growth (Figure 3B−D). We observed increased bioluminescence along time that correlates with the tumor size by tumor palpation (data not shown). The analysis of BxPC3-SELWP cells after disaggregation of the tumors generated in the transplanted mice showing a stable eGFP expression (Figure 3E,F), allowing the study of potential changes in the BxPC-3-SELWP phenotype after tumor generation. Interestingly, the expression of eGFP and NanoLuc in BxPC-3-SELWP cells did not significantly alter the expression pattern of both CD44 and CXCR4 PCSC markers (Figure 3G). In conclusion, we generated and validated a suitable pancreatic tumor model based on BxPC-3 PCC that stably expressed both eGFP and NanoLuc, which can be useful for *in vitro* phenotypic analysis and *in vivo* tracking.





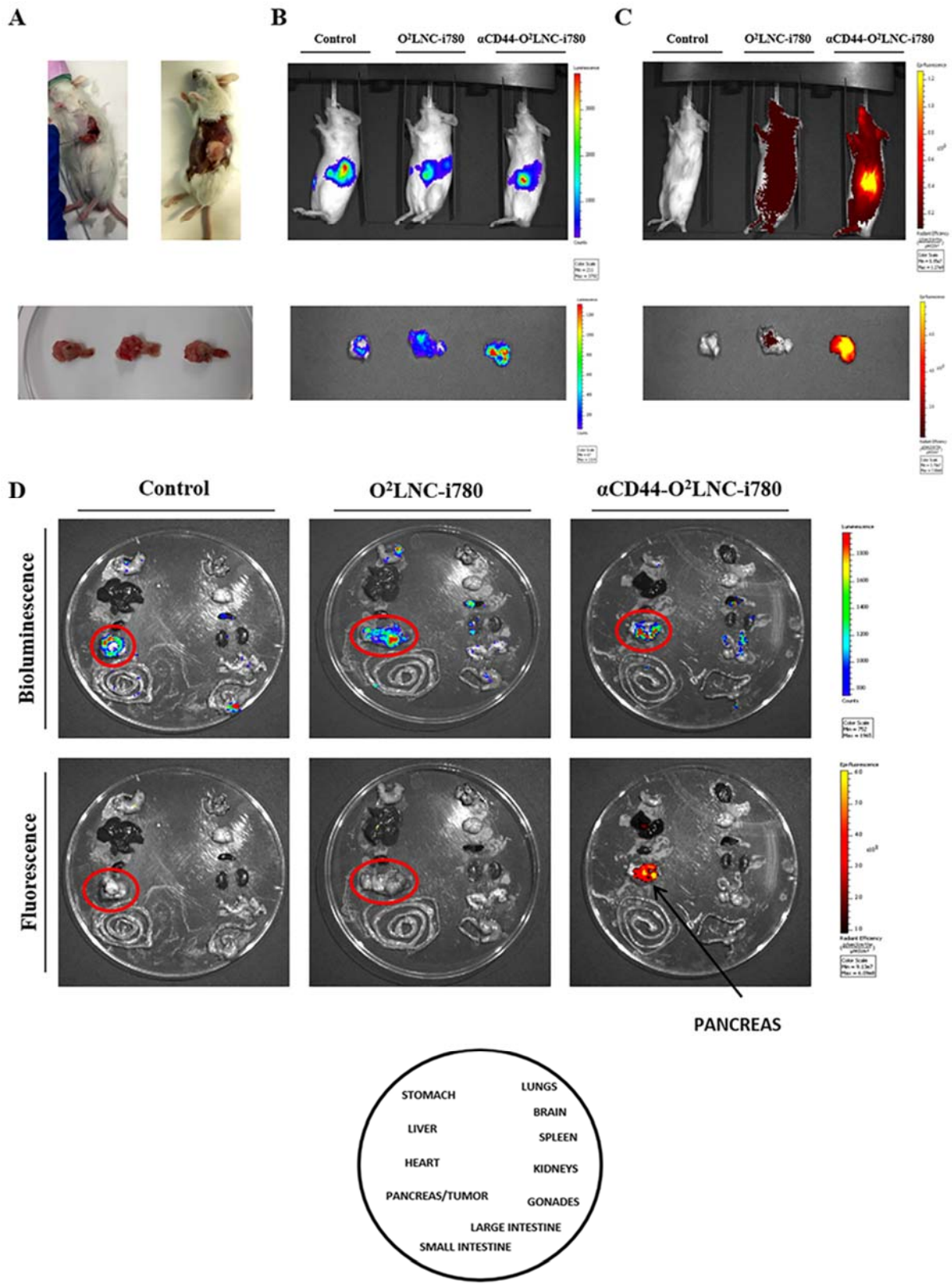

PANCREAS





**Figure 4.** *In vivo assay of O2LNC-i780 and αCD44-O2LNC-i780 targeting tumor. (A) Representative images of the orthotopic xenograft pancreatic tumor generation inoculating the cells (left image) and once the tumor was generated (right image); and excised tumor ex vivo (bottom image). (B) Representatives images of mice (top image) and tumors (bottom image) with Nano-Luc measured by bioluminescence. (C) Representatives images of mice and tumors treated with PBS, O2LNC-i780, and αCD44-O2LNC-i780 measured by fluorescence. (D) Representative images of the mice organs after treatment with PBS, O2LNC-i780, and αCD44-O2LNC-i780 measured by both bioluminescence and by fluorescence.*

**In vivo biodistribution and targeting assay.**

In order to verify that our functionalized αCD44-O2LNCs were specifically directed to the tumor area, an in vivo assay was performed. In this experiment, an orthotopic xenotransplant was performed in the pancreas of the mice, and the tumor was generated as described[75] but with a modification in the protocol by implanting PCSCs instead of patient tumor samples. BxPC-3 PCSCs transduced with the SELWPLVs were used to generate the tumor and the αCD44-O$^2$LNC were loaded with IR 780-iodide dye dissolved in the oily core. We choose this dye because it emits in the near infrared and can be easily detected by the in vivo imaging systems (IVIS), also avoiding background noise caused by the mouse's own autofluorescence. 76 When tumors reached a size of 0.5 cm3, 100 μL of O$^2$LNC-i780, and αCD44-O$^2$LNC-i780 were administered in the tail vein of the mice in a single dose. After one week, the mice were analyzed, euthanized, and the tumors were removed as well as all the organs. As shown in Figure 4A, 2 × 105 BxPC-3-SELWP PCSCs were able to engraft in the pancreases of mice only two weeks after inoculation and the growth was easily monitored by bioluminescence (Figure 4B). In addition, we were able to detect micrometastatic sites in organs close to the pancreas,





such as stomach, kidneys, or spleen among others (Figure 4D, top panel). An additional advantage of our dual vector is that the targeting properties of $O^2$LNC-i780 and αCD44-O2LNC-i780 were also visualized by fluorescence in the same mice. As can be observed in Figure 4C, O2LNC-i780s were distributed evenly throughout the mice one-week postinoculation. On the contrary, although the αCD44-$O^2$LNCi780s also showed a diffuse distribution in the mice , they strongly accumulated in the tumor area (Figure 4C). This fact was confirmed in the excised PCSC tumors in concordance with the signal intensity emitted by eGFP in the tumors (Figure 4B,C, down panel) as well as in the organs excised from the mice (Figure 4D). In agreement with other studies using targeted tumor nanoparticles[66,77], our olive oil-based nanosystem specifically targets PC tumors which could serve as a noninvasive imaging modality for molecular and cellular tracing and specific carrier of hydrophobic drugs against PCSCs. More specifically, Han et al. demonstrated the targeting ability of anti-CD326-grafted gadolinium ion-doped up conversion nanoparticle-based micelles on BxPC-3 PCCs overexpressing this transmembrane glycoprotein[78]. In addition to these studies, here, we show an enhanced targeting ability of our olive oil-based nanosystem on a PCSC orthotopic xenotransplant in vivo model.

**Conclusions**

$O^2$LNC functionalized with an anti-CD44 antibody and with specific recognition properties directed against PCSCs for the first time. The main advantages of this nanosystem are as follows: (i) the use of olive oil as a biocompatible and biodegradable essential component; (ii) a core−shell composition able to encapsulate hydrophobic drugs inside, and (iii) the ability to be efficiently coated with antibody molecules by means of a covalent binding. We demonstrated the successful anti-CD44 antibody surface functionalization through the modification and optimization of a simple and





reproducible ECDI covalence method. The complete physicochemical characterization of the immuno-NCs showed the accurate binding of the antibody molecules to the carboxyl surface groups of bare $O^2$LNCs. Additionally, the final surface hydrophilicity of the αCD44-$O^2$LNC favors the appearance of hydration repulsive forces allowing the nanosystem to remain colloidally stable in saline solutions and in typical cell culture media. Furthermore, the specific immunological recognition of the CD44 cell membrane receptors has been revealed in vitro, showing the adequate surface disposition of the CD44 antibody molecules immobilized on the $O^2$LNC.

Our in vitro results also supported the enhanced targeting activity and receptor-mediated binding mechanism of αCD44-$O^2$LNC with CD44 overexpressing PCSCs and the increased cytotoxicity of up to 4 times of PTX-loaded αCD44-$O^2$LNC on PCSCs in comparison to free PTX. This fact indicates the ability of our targeted nanosystem to increase the antitumor potency of PTX in cancer cells with stemness properties. Finally, both the efficient targeted delivery to PCSCs and the noninvasive imaging cell tracking of αCD44-$O^2$LNC has been demonstrated in vivo, which suggest their promising potential in targeted tumor synergistic theranostics. In summary, this nanosystem based on anti-CD44-functionalized $O^2$LNC might be a potent candidate for the treatment of CD44 overexpressing CSCs.

Finally, it is important to note that several modifications can be made to this novel nanosystem that might broaden the range of applications, such as carrying different drugs, the alternative use of other tracker molecules, depending on the detection technique, or the conjugation of other specific molecules, such as peptides or antibodies, for different tumors and even for different pathological areas. In the near future, further in vivo studies will be carried out to evaluate the versatility of this nanosystem in the antitumor effect of αCD44-$O^2$LNC loaded with several drugs.





## ASSOCIATED CONTENT

## Supporting information:

$^1$H NMR spectrum of intact O$^2$LNC in H$_2$O−D$_2$O, $^1$H NMR spectrum (expanded) of intact O$^2$LNC in H$_2$O−D$_2$O, comparison of $^1$H NMR spectra of intact O$^2$LNC and olive oil, diffusion NMR experiment with intact O$^2$LNC, comparison of $^1$H NMR spectra of intact O$^2$LNC and Pluronic F-68, $^1$H NMR spectrum of disintegrated O$^2$LNC in CDCl$_3$, comparison of $^1$H NMR spectra (CDCl$_3$) of disintegrated O$^2$LNC and olive oil, O$^2$LNC and Epikuron 145V, O$^2$LNC and Pluronic F-68, and O$^2$LNC and DCA, $^1$H NMR spectrum of disintegrated O$^2$LNC in CD$_3$OD, comparison of $^1$H NMR spectra (CD$_3$OD) of disintegrated O$^2$LNC and olive oil, O$^2$LNC and Epikuron 145V, O$^2$LNC and Pluronic F-68, and O$^2$LNC and DCA, size distribution histograms by intensity of O$^2$LNCs and αCD44-O$^2$LNCs in different media, representative control DOT PLOTS of BxPC-3 PCCs and PCSCs analyzed by flow cytometry, and cytotoxic effect of free PTX, O$^2$LNC-PTX, and αCD44-O$^2$LNC-PTX on BxPC-3 PCCs and PCSCs (PDF)

## Acknowledgments

This work was supported by grants from the Ministry of Economy and Competitiveness (MINECO, National Program for Research Aimed at the Challenges of Society, project reference MAT2015-63644-C2-2-R & MAT2015-63644-C2-1-R FEDER funds), by the Ministerio de Ciencia, Innovación y Universidades grant numbers RTI2018-101309-B-C22 & RTI2018-101309-B-C21 (FEDER Funds), by the Consejería de Economía, Conocimiento, Empresas y Universidad de la Junta de Andalucía (SOMM17/6109/UGR, FEDER Funds),by the Instituto de Salud Carlos III, Spain (www.isciii.es) and Fondo Europeo de Desarrollo Regional (FEDER), from the





European Union, through the research grants PI15/02015 and PI18/00337 and by Consejería de Salud de la Junta de Andalucía through the research salud-2016000073332-TRA to FM and by the Chair "Doctors Galera-Requena in cancer stem cell research". S.A.N.M. and M.T.M. acknowledges the MINECO for providing a PhD fellowship (FPI and FPU respectively) through the project MAT2015-63644-C2-2-R and FPU16/05467. The NMR spectrometer used in this work were purchased via a grant for scientific and technological infrastructure from the Ministerio de Ciencia e Innovación [Grant No. PCT-010000-2010-4).

## AUTHOR INFORMATION

### Corresponding Author

* E-mail: jmarchal@go.ugr.es

* E-mail: jmpeula@uma.es

### ORCID

Juan Antonio Marchal Corrales: https://orcid.org/0000-0002-4996-8261

José Manuel Peula García: https://orcid.org/0000-0002-0869-9579

### Notes

The authors declare no competing financial interest






**REFERENCES**

(1) Duncan, R. Nanomedicines in Action. Pharm. J. 2004, 273, 485−488.

(2) Sorg, C. Scientific Forward Look on Nanomedicine. Eur. Sci.Found. Policy Brief. 2005, 23, 1−6.

(3) Ferrari, M. Cancer Nanotechnology: Opportunities and Challenges. Nat. Rev. Cancer 2005, 5, 161−171.

(4) Zhukov, N. V.; Tjulandin, S. A. Targeted Therapy in the Treatment of Solid Tumors: Practice Contradicts Theory. Biochem 2008, 73, 605−618.

(5) Gu, F.; Zhang, L.; Teply, B. A.; Mann, N.; Wang, A.; Radovic-Moreno, A. F.; Langer, R.; Farokhzad, O. C. Precise Engineering of Targeted Nanoparticles by Using Self-Assembled Biointegrated Block Copolymers. Proc. Natl. Acad. Sci. U.S.A. 2008, 105, 2586−2591.

(6) Heath, J. R.; Davis, M. E. Nanotechnology and Cancer. Annu. Rev. Med. 2008, 59, 251−265.

(7) Peer, D.; Karp, J. M.; Hong, S.; Farokhzad, O. C.; Margalit, R.; Langer, R. Nanocarriers as an emerging platform for cancer therapy. Nat. Nanotechnol. 2007, 2, 751−760.

(8) Delogu, L. G.; Venturelli, E.; Manetti, R.; Pinna, G. A.; Carru, C.; Madeddu, R.; Murgia, L.; Sgarrella, F.; Dumortier, H.; Bianco, A. Ex Vivo Impact of Functionalized Carbon Nanotubes on Human Immune Cells. Nanomedicine 2012, 7, 231−243.

(9) Pescatori, M.; Bedognetti, D.; Venturelli, E.; Ménard-Moyon, C.; Bernardini, C.;







Muresu, E.; Piana, A.; Maida, G.; Manetti, R.; Sgarrella, F.; et al. Functionalized Carbon Nanotubes as Immunomodulator Systems. Biomaterials 2013, 34, 4395−4403.

(10) Yoo, H. S.; Park, T. G. Biodegradable Polymeric Micelles Composed of Doxorubicin Conjugated PLGA-PEG Block Copolymer. J. Controlled Release 2001, 70, 63−70.

(11) Allen, T. M. Ligand-Targeted Therapeutics in Anticancer Therapy. Nat. Rev. Cancer 2002, 2, 750−763.

(12) Torchilin, V. P. Recent Advances with Liposomes as Pharmaceutical Carriers. Nat. Rev. Drug Discovery 2005, 4, 145−160.

(13) Thierry, B. Drug Nanocarriers and Functional Nanoparticles: Applications in Cancer Therapy. Curr. Drug Delivery 2009, 6, 391−403.

(14) Li, J.; Chen, Y.; Zeng, L.; Lian, G.; Chen, S.; Li, Y.; Yang, K.; Huang, K. A Nanoparticle Carrier for Co-Delivery of Gemcitabine and Small Interfering Rna in Pancreatic Cancer Therapy. J. Biomed. Nanotechnol. 2016, 12, 1654−1666.

(15) Li, Y.; Chen, Y.; Li, J.; Zhang, Z.; Huang, C.; Lian, G.; Yang, K.; Chen, S.; Lin, Y.; Wang, L.; et al. Co-Delivery of MicroRNA-21 Antisense Oligonucleotides and Gemcitabine Using Nanomedicine for Pancreatic Cancer Therapy. Cancer Sci. 2017, 108, 1493−1503.

(16) Skandalis, S. S.; Gialeli, C.; Theocharis, A. D.; Karamanos, N. K. Advances and Advantages of Nanomedicine in the Pharmacological Targeting of Hyaluronan-CD44 Interactions and Signaling in Cancer; Academic Press Inc., 2014; Vol. 123, pp 277−317.







(17) Yang, Y.; Zhao, X.; Li, X.; Yan, Z.; Liu, Z.; Li, Y. Effects of Anti-CD44 Monoclonal Antibody IM7 Carried with Chitosan Polylactic Acid-Coated Nano-Particles on the Treatment of Ovarian Cancer. Oncol. Lett. 2017, 13, 99−104.

(18) Platt, V. M.; Szoka, F. C. Anticancer Therapeutics: Targeting Macromolecules and Nanocarriers to Hyaluronan or CD44, a Hyaluronan Receptor. Mol. Pharm. 2008, 5, 474−486.

(19) Moghimi, S. M.; Hunter, A. C.; Murray, J. C. Long-Circulating and Target-Specific Nanoparticles: Theory to Practice. Pharmacol. Rev. 2001, 53, 283−318.

(20) Stockler, M.; Wilcken, N. R.; Ghersi, D.; Simes, R. J. Systematic Reviews of Chemotherapy and Endocrine Therapy in Metastatic Breast Cancer. Cancer Treatment Reviews; W.B. Saunders Ltd, 2000; pp 151−168.

(21) Naor, D.; Wallach-Dayan, S. B.; Zahalka, M. A.; Sionov, R. V. Involvement of CD44, a Molecule with a Thousand Faces, in Cancer Dissemination. Semin. Cancer Biol. 2008, 18, 260−267.

(22) Naor, D.; Nedvetzki, S.; Golan, I.; Melnik, L.; Faitelson, Y. CD44 in Cancer. Crit. Rev. Clin. Lab. Sci. 2002, 39, 527−579.

(23) Thapa, R.; Wilson, G. D. The Importance of CD44 as a Stem Cell Biomarker and Therapeutic Target in Cancer. Stem Cells Int. 2016, 2016, 2087204.

(24) Yan, Y.; Zuo, X.; Wei, D. Concise Review: Emerging Role of CD44 in Cancer Stem Cells: A Promising Biomarker and Therapeutic Target. Stem Cells Transl. Med. 2015, 4, 1033−1043.

(25) Bray, F.; Ferlay, J.; Soerjomataram, I.; Siegel, R. L.; Torre, L. A.; Jemal, A. Global








Cancer Statistics 2018: GLOBOCAN Estimates of Incidence and Mortality Worldwide for 36 Cancers in 185 Countries. Ca-Cancer J. Clin. 2018, 68, 394−424.

(26) Boulaiz, H.; Ramos, M. C.; Griñán-Lisón, C.; García-Rubiño, M. E.; Vicente, F.; Marchal, J. A. What's new in the diagnosis of pancreatic cancer: a patent review (2011-present). Expert Opin. Ther. Pat. 2017, 27, 1319−1328.

(27) Ramos, M. C.; Boulaiz, H.; Griñan-Lison, C.; Marchal, J. A.; Vicente, F. What's New in Treatment of Pancreatic Cancer: A Patent Review (2010−2017). Expert Opin. Ther. Pat. 2017, 27, 1251−1266.

(28) Hernández-Camarero, P.; Jiménez, G.; López-Ruiz, E.; Barungi, S.; Marchal, J. A.; Perán, M. Revisiting the Dynamic Cancer Stem Cell Model: Importance of Tumour Edges. Crit. Rev. Oncol. Hematol. 2018, 131, 35−45.

(29) Sánchez-Moreno, P.; Ortega-Vinuesa, J. L.; Boulaiz, H.; Marchal, J. A.; Peula-García, J. M. Synthesis and Characterization of Lipid Immuno-Nanocapsules for Directed Drug Delivery: Selective Antitumor Activity against HER2 Positive Breast-Cancer Cells. Biomacromolecules 2013, 14, 4248−4259.

(30) Zhou, H.; Yu, W.; Guo, X.; Liu, X.; Li, N.; Zhang, Y.; Ma, X. Synthesis and Characterization of Amphiphilic Glycidol−Chitosan−Deoxycholic Acid Nanoparticles as a Drug Carrier for Doxorubicin. Biomacromolecules 2010, 11, 3480−3486.

(31) Samstein, R. M.; Perica, K.; Balderrama, F.; Look, M.; Fahmy, T. M. The Use of Deoxycholic Acid to Enhance the Oral Bioavailability of Biodegradable






Nanoparticles. Biomaterials 2008, 29, 703−708.

(32) Salvati, A.; Pitek, A. S.; Monopoli, M. P.; Prapainop, K.; Bombelli, F. B.; Hristov, D. R.; Kelly, P. M.; Åberg, C.; Mahon, E.; Dawson, K. A. Transferrin-Functionalized Nanoparticles Lose Their Targeting Capabilities When a Biomolecule Corona Adsorbs on the Surface. Nat. Nanotechnol. 2013, 8, 137−143.

(33) Sánchez-Moreno, P.; Ortega-Vinuesa, J. L.; Martín-Rodríguez, A.; Boulaiz, H.; Marchal-Corrales, J. A.; Peula-García, J. M. Characterization of Different Functionalized Lipidic Nanocapsules as Potential Drug Carriers. Int. J. Mol. Sci. 2012, 13, 2405−2424.

(34) Jiménez, G.; Hackenberg, M.; Catalina, P.; Boulaiz, H.; Griñán- Lisón, C.; García, M. Á.; Perán, M.; López-Ruiz, E.; Ramírez, A.; Morata-Tarifa, C.; et al. Mesenchymal Stem Cell's Secretome Promotes Selective Enrichment of Cancer Stem-like Cells with Specific Cytogenetic Profile. Cancer Lett. 2018, 429, 78−88.

(35) Calvo, P.; Remuñán-López, C.; Vila-Jato, J. L.; Alonso, M. J. Novel Hydrophilic Chitosan-Polyethylene Oxide Nanoparticles as Protein Carriers. J. Appl. Polym. Sci. 1997, 63, 125−132.

(36) Delgado-Calvo-Flores, J. M.; Peula-García, J. M.; Martínez-García, R.; Callejas-Fernández, J. Experimental Evidence of Hydration Forces between Polymer Colloids Obtained by Photon Correlation Spectroscopy Measurements. J. Colloid Interface Sci. 1997, 189, 58−65.

(37) Demaison, C.; Parsley, K.; Brouns, G.; Scherr, M.; Battmer, K.; Kinnon, C.; Grez,









M.; Thrasher, A. J. High-Level Transduction and Gene Expression in Hematopoietic Repopulating Cells Using a Human Imunodeficiency Virus Type 1-Based Lentiviral Vector Containing an Internal Spleen Focus Forming Virus Promoter. Hum. Gene Ther. 2002, 13, 803−813.

(38) Benabdellah, K.; Gutierrez-Guerrero, A.; Cobo, M.; Muñoz, P.; Martín, F. A Chimeric HS4-SAR Insulator (IS2) That Prevents Silencing and Enhances Expression of Lentiviral Vectors in Pluripotent Stem Cells. PLoS One 2014, 9, No. e84268.

(39) Urbaniak, T.; Musiał, W. Influence of Solvent Evaporation Technique Parameters on Diameter of Submicron Lamivudine-Poly- ε-Caprolactone Conjugate Particles. Nanomaterials 2019, 9, 1240.

(40) Díaz-Torres, R.; López-Arellano, R.; Escobar-Chávez, J. J.; García-García, E.; Domínguez-Delgado, C. L.; Ramírez-Noguera, P. Effect of Size and Functionalization of Pharmaceutical Nanoparticles and Their Interaction with Biological Systems. Handbook of Nanoparticles; Springer International Publishing, 2015; pp 1041−1060.

(41) Panyam, J.; Labhasetwar, V. Biodegradable Nanoparticles for Drug and Gene Delivery to Cells and Tissue. Adv. Drug Delivery Rev. 2003, 55, 329−347.

(42) Garcia-Fuentes, M.; Torres, D.; Martín-Pastor, M.; Alonso, M. J. Application of NMR Spectroscopy to the Characterization of PEGStabilized Lipid Nanoparticles. Langmuir 2004, 20, 8839−8845.

(43) Mammen, M.; Choi, S.-K.; Whitesides, G. M. Polyvalent Interactions in Biological Systems: Implications for Design and Use of Multivalent Ligands and Inhibitors.







Angew. Chem., Int. Ed. 1998, 37, 2754−2794.

(44) Mukherjee, S.; Ghosh, R. N.; Maxfield, F. R. Endocytosis. Physiological Reviews; American Physiological Society, 1997; pp 759− 803.

(45) Weissleder, R.; Kelly, K.; Sun, E. Y.; Shtatland, T.; Josephson, L. Cell-Specific Targeting of Nanoparticles by Multivalent Attachment of Small Molecules. Nat. Biotechnol. 2005, 23, 1418−1423.

(46) Stefanick, J. F.; Ashley, J. D.; Kiziltepe, T.; Bilgicer, B. A Systematic Analysis of Peptide Linker Length and Liposomal Polyethylene Glycol Coating on Cellular Uptake of Peptide-Targeted Liposomes. ACS Nano 2013, 7, 2935−2947.

(47) Elias, D. R.; Poloukhtine, A.; Popik, V.; Tsourkas, A. Effect of Ligand Density, Receptor Density, and Nanoparticle Size on Cell Targeting. Nanomedicine 2013, 9, 194−201.

(48) Peula, J. M.; Hidalgo-Alvarez, R.; De Las Nieves, F. J. Covalent Binding of Proteins to Acetal-Functionalized Latexes. I. Physics and Chemical Adsorption and Electrokinetic Characterization. J. Colloid Interface Sci. 1998, 201, 132−138.

(49) Goldstein, D.; Gofrit, O.; Nyska, A.; Benita, S. Anti-HER2 Cationic Immunoemulsion as a Potential Targeted Drug Delivery System for the Treatment of Prostate Cancer. Cancer Res. 2007, 67, 269−275.

(50) Nebija, D.; Kopelent-Frank, H.; Urban, E.; Noe, C. R.; Lachmann, B. Comparison of Two-Dimensional Gel Electrophoresis Patterns and MALDI-TOF MS Analysis of Therapeutic Recombinant Biomacromolecules Monoclonal Antibodies Trastuzumab and Rituximab. J. Pharm. Biomed. Anal. 2011, 56, 684−691.









(51) Pop-Georgievski, O.; Popelka, Š.; Houska, M.; Chvostová, D.; Proks, V.; Rypáček, F. Poly(Ethylene Oxide) Layers Grafted to Dopamine-Melanin Anchoring Layer: Stability and Resistance to Protein Adsorption. Biomacromolecules 2011, 12, 3232−3242.

(52) Delcroix, M. F.; Huet, G. L.; Conard, T.; Demoustier-Champagne, S.; Du Prez, F. E.; Landoulsi, J.; Dupont-Gillain, C. C. Design of Mixed PEO/PAA Brushes with Switchable Properties toward Protein Adsorption. Biomacromolecules 2013, 14, 215−225.

(53) Torcello-Gómez, A.; Santander-Ortega, M. J.; Peula-García, J. M.; Maldonado-Valderrama, J.; Gálvez-Ruiz, M. J.; Ortega-Vinuesa, J.L.; Martín-Rodríguez, A. Adsorption of Antibody onto Pluronic F68-Covered Nanoparticles: Link with Surface Properties. Soft Matter 2011, 7, 8450−8461.

(54) Martin, A.; Puig, J.; Galisteo, F.; Serra, J.; Hidalgo-Alvarez, R. On Some Aspect of the Adsorption of Immunoglobulin-G Molecules on Polystyrene Microspheres. J. Dispersion Sci. Technol. 1992, 13, 399−416.

(55) Peula, J. M.; Hidalgo-Alvarez, R.; De Las Nieves, F. J. Coadsorption of IgG and BSA onto Sulfonated Polystyrene Latex: I. Sequential and Competitive Coadsorption Isotherms. J. Biomater. Sci., Polym. Ed. 1996, 7, 231−240.

(56) Peula-Garcia, J. M.; Hidaldo-Alvarez, R.; De las Nieves, F. J.Protein Co-Adsorption on Different Polystyrene Latexes: Electrokinetic Characterization and Colloidal Stability. Colloid Polym. Sci. 1997, 275, 198−202.

(57) Molina-Bolívar, J. A.; Ortega-Vinuesa, J. L. How Proteins Stabilize Colloidal Particles by Means of Hydration Forces. Langmuir 1999, 15, 2644−2653.







(58) Santander-Ortega, M. J.; Lozano-López, M. V.; Bastos-González, D.; Peula-García, J. M.; Ortega-Vinuesa, J. L. Novel Core-Shell Lipid-Chitosan and Lipid-Poloxamer Nanocapsules: Stability by Hydration Forces. Colloid Polym. Sci. 2010, 288, 159−172.

(59) Farace, C.; Sánchez-Moreno, P.; Orecchioni, M.; Manetti, R.; Sgarrella, F.; Asara, Y.; Peula-García, J. M.; Marchal, J. A.; Madeddu, R.; Delogu, L. G. Immune Cell Impact of Three Differently Coated Lipid Nanocapsules: Pluronic, Chitosan and Polyethylene Glycol. Sci. Rep. 2016, 6, 18423.

(60) Xu, Z.; Jia, Y.; Huang, X.; Feng, N.; Li, Y. Rapid Induction of Pancreatic Cancer Cells to Cancer Stem Cells via Heterochromatin Modulation. Cell Cycle 2018, 17, 1487−1495.

(61) Ning, X.; Du, Y.; Ben, Q.; Huang, L.; He, X.; Gong, Y.; Gao, J.; Wu, H.; Man, X.; Jin, J.; et al. Bulk Pancreatic Cancer Cells Can Convert into Cancer Stem Cells(CSCs) in Vitro and 2 Compounds Can Target These CSCs. Cell Cycle 2016, 15, 403−412.

(62) Hernández-Camarero, P.; López-Ruiz, E.; Griñán-Lisón, C.; García, M. Á.; Chocarro-Wrona, C.; Marchal, J. A.; Kenyon, J.; Perán, M. Pancreatic (pro)Enzymes Treatment Suppresses BXPC-3 Pancreatic Cancer Stem Cell Subpopulation and Impairs Tumour Engrafting. Sci. Rep. 2019, 9, 11359.

(63) Rejman, J.; Oberle, V.; Zuhorn, I. S.; Hoekstra, D. Sizedependent internalization of particles via the pathways of clathrin- and caveolae-mediated endocytosis. Biochem. J. 2004, 377, 159−169.

(64) Kesharwani, P.; Banerjee, S.; Padhye, S.; Sarkar, F. H.; Iyer, A. K. Hyaluronic







Acid Engineered Nanomicelles Loaded with 3,4-Difluorobenzylidene Curcumin for Targeted Killing of CD44+ Stem-Like Pancreatic Cancer Cells. Biomacromolecules 2015, 16, 3042−3053.

(65) Qian, C.; Wang, Y.; Chen, Y.; Zeng, L.; Zhang, Q.; Shuai, X.;Huang, K. Suppression of Pancreatic Tumor Growth by Targeted Arsenic Delivery with Anti-CD44v6 Single Chain Antibody Conjugated Nanoparticles. Biomaterials 2013, 34, 6175−6184.

(66) Cano-Cortes, M. V.; Navarro-Marchal, S. A.; Ruiz-Blas, M. P.; Diaz-Mochon, J. J.; Marchal, J. A.; Sanchez-Martin, R. M. A Versatile Theranostic Nanodevice Based on an Orthogonal Bioconjugation Strategy for Efficient Targeted Treatment and Monitoring of Triple Negative Breast Cancer. Nanomedicine 2020, 24, 102−120.

(67) Trabulo, S.; Aires, A.; Aicher, A.; Heeschen, C.; Cortajarena, A. L. Multifunctionalized Iron Oxide Nanoparticles for Selective Targeting of Pancreatic Cancer Cells. Biochim. Biophys. Acta, Gen. Subj. 2017, 1861, 1597−1605.

(68) Liebmann, J.; Cook, J.; Lipschultz, C.; Teague, D.; Fisher, J.;Mitchell, J. Cytotoxic studies of paclitaxel (Taxol) in human tumour cell lines. Br. J. Cancer 1993, 68, 1104−1109.

(69) Okamoto, Y.; Taguchi, K.; Sakuragi, M.; Imoto, S.; Yamasaki, K.; Otagiri, M. Preparation, Characterization, and in Vitro/in Vivo Evaluation of Paclitaxel-Bound Albumin-Encapsulated Liposomes for the Treatment of Pancreatic Cancer. ACS Omega 2019, 4, 86938700.







(70) Wu, S.-t.; Fowler, A. J.; Garmon, C. B.; Fessler, A. B.; Ogle, J. D.; Grover, K. R.; Allen, B. C.; Williams, C. D.; Zhou, R.; Yazdanifar, M.; et al. Treatment of Pancreatic Ductal Adenocarcinoma with Tumor Antigen Specific-Targeted Delivery of Paclitaxel Loaded PLGA Nanoparticles. BMC Cancer 2018, 18, 457.

(71) Dubey, N.; Shukla, J.; Hazari, P. P.; Varshney, R.; Ganeshpurkar, A.; Mishra, A. K.; Trivedi, P.; Bandopadhaya, G. P. Preparation and Biological Evaluation of Paclitaxel Loaded Biodegradable PCL/PEG Nanoparticles for the Treatment of Human Neuroendocrine Pancreatic Tumor in Mice. Hell. J. Nucl. Med. 2012, 15, 9−15.

(72) Lee, C. J.; Spalding, A. C.; Ben-Josef, E.; Wang, L.; Simeone, D. M. In Vivo Bioluminescent Imaging of Irradiated Orthotopic Pancreatic Cancer Xenografts in Nonobese Diabetic-Severe Combined Immunodeficient Mice: A Novel Met Targeting and Assaying Efficacy of Ionizing Radiation. Transl. Oncol. 2010, 3, 153−159.

(73) Stacer, A. C.; Nyati, S.; Moudgil, P.; Iyengar, R.; Luker, K. E.; Rehemtulla, A.; Luker, G. D. NanoLuc Reporter for Dual Luciferase Imaging in Living Animals. Mol. Imag. 2013, 12, 1−13.

(74) Luker, C. G.; Ehlerding, E. B.; Cai, W. NanoLuc: A Small Luciferase Is Brightening Up the Field of Bioluminescence. Bioconjugate Chem. 2016, 27, 1175−1187.

(75) Rubio-Viqueira, B.; Jimeno, A.; Cusatis, G.; Zhang, X.; Iacobuzio-Donahue, C.; Karikari, C.; Shi, C.; Danenberg, K.; Danenberg, P. V.; Kuramochi, H.; et al. An in Vivo Platform for Translational Drug Development in Pancreatic Cancer. Clin. Cancer Res. 2006, 12, 4652−4661.







(76) Yi, X.; Zhang, J.; Yan, F.; Lu, Z.; Huang, J.; Pan, C.; Yuan, J.; Zheng, W.; Zhang, K.; Wei, D.; et al. Synthesis of IR-780 Dye-Conjugated Abiraterone for Prostate Cancer Imaging and Therapy. Internet J. Oncol. 2016, 49, 1911−1920.

(77) Miller, A. D. Lipid-Based Nanoparticles in Cancer Diagnosis and Therapy. J. Drug Delivery 2013, 2013, 165981.

(78) Han, Y.; An, Y.; Jia, G.; Wang, X.; He, C.; Ding, Y.; Tang, Q. Facile Assembly of Upconversion Nanoparticle-Based Micelles for Active Targeted Dual-Mode Imaging in Pancreatic Cancer. J. Nanobiotechnol. 2018, 16, 7.